\documentclass[10pt,conference,letterpaper,final]{IEEEtran}
\IEEEoverridecommandlockouts

\usepackage{graphicx}
\usepackage[table]{xcolor}
\usepackage{tikz}
\usepackage[utf8]{inputenc} 
\usepackage[T1]{fontenc}
\usepackage{amsmath}
\usepackage{amssymb}
\usepackage{amsfonts} 
\usepackage[normalem]{ulem}
\usepackage{bbm}
\usepackage{url}
\usepackage{makecell} 
\usepackage{xstring}

\usepackage{algorithm}
\usepackage[noend]{algpseudocode}
\algrenewcommand\algorithmicrequire{\textbf{Input:}}
\algrenewcommand\algorithmicensure{\textbf{Output:}}
\algrenewcommand{\algorithmiccomment}[1]{\hfill{\color{blue}$\triangleright$#1}}

\usepackage{siunitx}
\usepackage{mathtools,zref-savepos}

\usepackage{float}
\usepackage[caption=false,font=footnotesize]{subfig}
\usepackage{booktabs}  
\usepackage{tabularx}  
\usepackage{multirow}
\usepackage{xspace}

\usepackage{ifdraft}
\ifoptionfinal{\usepackage[final]{fixme}}{\usepackage[draft]{fixme}}
\fxsetup{mode=multiuser,theme=color,layout=inline,nomargin} 
\FXRegisterAuthor{DL}{dl}{\color{teal}\fbox{DL}}
\FXRegisterAuthor{DB}{db}{\color{blue}\fbox{DB}}
\FXRegisterAuthor{OK}{ok}{\color{olive}\fbox{OK}}
\FXRegisterAuthor{GY}{gy}{\color{purple}\fbox{GY}}

\newcommand{\DL}[1]{\DLnote{\color{teal}#1}}
\newcommand{\DB}[1]{\DBnote{\color{blue}#1}}
\newcommand{\OK}[1]{\OKnote{\color{olive}#1}}
\newcommand{\GY}[1]{\GYnote{\color{purple}#1}}

\usepackage{balance}

\usepackage{xspace}
\usepackage{xparse}
\usepackage{enumitem}

\usepackage{comment}

\usepackage[hidelinks]{hyperref}

\usepackage[capitalise]{cleveref}
\crefname{equation}{}{}
\crefname{section}{Sec.}{Secs.}

\newcommand{\Set}[1]{\ensuremath{\mathcal{#1}}\xspace}
\newcommand{\Shortstack}[2][t]{\begin{tabular}[#1]{@{}l@{}}#2\end{tabular}}
\newcommand{\range}[1]{\ensuremath[#1]}

\DeclareSIUnit{\million}{\text{million}}

\crefname{equation}{}{}
\Crefname{equation}{Eq.}{Eqs.}
\crefname{alt}{Alternative}{Alternatives}
\Crefname{emb}{Embedding}{Embeddings}
\crefname{emb}{}{Embeddings}
\newcommand{\eref}[1]{Eq.~\eqref{#1}}

\crefname{define}{Definition}{Definitions}

\crefname{observation}{Observation}{Observations}

\newcommand{\dc}{datacenter\xspace}
\newcommand{\dcs}{datacenters\xspace}

\newcommand\MyIncludeGraphics[2][]{
    \IfFileExists{#2}{%
        \includegraphics[#1]{#2}%
    }{%
        \missingfigure[figwidth=7.0cm]{Missing #2}%
    }%
}%

\newcommand{\T}[1]{\par\noindent\rule{0pt}{\baselineskip}\textit{#1.}}

\newcommand{\set}[1]{\left\{#1\right\}}         

\newcommand{\abs}[1]{\left\vert#1\right\vert}

\DeclareMathOperator{\concat}{+\kern -0.4em+}

\newcommand{\newVar}[2]{\newcommand{#1}{\ensuremath{#2}\xspace}}
  \newVar{\server}{S}
  \newVar{\client}{C}
  \newVar{\rclient}{R_c}
  \newVar{\rserver}{R_s}

\providecommand{\eg}{\emph{e.g.,} }

\newcommand{\ignore}[1]{}
\newcommand{\Small}{\footnotesize} 
\newcommand{\blue}[2][blue]{{\color{#1}#2}}
\newcommand{\Rplus}{\mathbb{R}^+}  
\newcommand{\Ldots}{\ifmmode\mathinner{\ldotp\kern-0.2em\ldotp\kern-0.2em\ldotp}\else.\kern-0.13em.\kern-0.13em.\fi}

\Crefname{problem}{Problem}{Problems}
\crefname{problem}{Problem}{Problems}

\newenvironment{problem}[2]{%
\vspace{-.3\baselineskip}
\phantomsection
\par\noindent\rule[0\baselineskip]{\linewidth}{0.5pt}
\raisebox{0.2\baselineskip}{\textbf{#1 (#2)}}
\par\noindent\rule[.6\baselineskip]{\linewidth}{0.1pt}
\footnotesize\\[-.7\baselineskip]\noindent}{%
\par\vspace{-.2\baselineskip}\rule[\baselineskip]{\linewidth}{0.5pt}
\vspace{-\baselineskip}
}

\newcommand{\PLAN}{\textsc{PLAN}\xspace}

\newcommand{\CVNE}{Virtual Network Embedding\xspace}
\newcommand{\VNE}{virtual network embedding\xspace}

\NewDocumentCommand\vne{o}
    {\textsc{vne}%
    \IfNoValueTF{#1}
        {\xspace}
        {(#1)\xspace}}
\crefformat{vne}{#2\vne{}#3}
\Crefformat{vne}{\Vne}

\newcommand{\CPvne}{Planned \CVNE}
\newcommand{\Pvne}{planned \VNE}
\NewDocumentCommand\pvne{o}
    {\textsc{Plan-\vne}%
    \IfNoValueTF{#1}
        {\xspace}
        {(#1)\xspace}}
\crefformat{pvne}{#2\pvne{}#3}
\Crefformat{pvne}{\Pvne}

\newcommand{\COfvne}{Offline \CVNE}
\newcommand{\Ofvne}{offline \VNE}
\NewDocumentCommand\ofvne{o}
    {\textsc{off-vne}%
    \IfNoValueTF{#1}
        {\xspace}
        {(#1)\xspace}}
\crefformat{ofvne}{#2\ofvne{}#3}
\Crefformat{ofvne}{\ofvne}

\NewDocumentCommand\genvne{o}
    {\textsc{gen-vne}%
    \IfNoValueTF{#1}
        {\xspace}
        {(#1)\xspace}}
\crefformat{genvne}{#2\genvne{}#3}
\Crefformat{genvne}{\genvne}

\newcommand{\COnvne}{Online \CVNE}
\newcommand{\Onvne}{online \VNE}
\NewDocumentCommand\onvne{o}
    {\textsc{on-vne}%
    \IfNoValueTF{#1}
        {\xspace}
        {(#1)\xspace}}
\crefformat{onvne}{#2\onvne{}#3}
\Crefformat{onvne}{\onvne}

\NewDocumentCommand\avne{o}
    {\textsc{agg-\vne}%
    \IfNoValueTF{#1}
        {\xspace}
        {(#1)\xspace}}
\crefformat{avne}{#2\avne{}#3}
\Crefformat{avne}{\Ovne}

\crefname{equation}{}{Eqs.}
\Crefname{equation}{Equation}{Equations}

\crefname{greedy}{}{}
\creflabelformat{greedy}{#2\textsc{\textbf{G\scalebox{.8}{\kern-0.1em reedy}}}#3}
\crefname{olive}{}{}
\creflabelformat{olive}{#2\textsc{\textbf{O\scalebox{.8}{\kern-0.1em live}}}#3}

\crefname{embed}{}{}
\creflabelformat{embed}{#2\textbf{Embed}#3}
\crefname{algorithm}{Algorithm}{Algorithms}

\setlist[description]{leftmargin=*}

\newlist{doforall}{enumerate*}{1}
\setlist[doforall]{label=\textsuperscript{\alph*}}

\newcommand{\myop}[1]{\operatorname{#1}}
\newcommand{\capacity}[1]{\myop{cap}(#1)}
\NewDocumentCommand\cost{om}
    {\IfNoValueTF{#1}
        {\myop{cost}(#2)}
        {\myop{cost}_{#1}(#2)}}

\newcommand{\graphsym}{\boldsymbol{G}}
\newcommand{\sgraphsym}{\boldsymbol{S}}

\NewDocumentCommand\gnet{d<>}
    {\IfNoValueTF{#1}
        {\sgraphsym}
        {\sgraphsym(#1)}}

\newcommand{\nnet}{v}
\newcommand{\snet}{s}
\NewDocumentCommand\pnet{o}{\boldsymbol{p}\IfNoValueTF{#1}{}{_{#1}}}
\newcommand{\lnet}[1][vw]{(#1)}

\NewDocumentCommand\gapp{d<>O{a}}
    {\IfNoValueTF{#1}
        {\graphsym_{#2}}
        {\graphsym(#1)_{#2}}}

\NewDocumentCommand\napp{d<>O{i}}
    {{#2}\IfNoValueTF{#1}
        {}
        {_{#1}}}
\NewDocumentCommand\lapp{d<>O{ij}}
    {(#2)\IfNoValueTF{#1}
        {}
        {_{#1}}}
\NewDocumentCommand\user{d<>}
    {\theta\IfNoValueTF{#1}
        {}
        {_{#1}}}
\newcommand{\sapp}{q}
\newcommand{\allapps}{\Set A}
\newcommand{\app}{a}
\newcommand{\allt}{\Set T}

\newcommand{\dapp}[1]{{\beta}^{#1}}
\newcommand{\dnet}[2]{{\eta}^{#1}_{#2}}

\NewDocumentCommand\reqs{o}
    {\IfNoValueTF{#1}
        {\Set R}
        {\Set R_{#1}}}
\NewDocumentCommand\rreqs{o}    
    {\IfNoValueTF{#1}
        {\tilde{\Set R}}
        {\tilde{\Set R_{#1}}}}
\newcommand{\donereqs}{\ensuremath{\reqs[\textsc{done}]}\xspace}
\newcommand{\planreqs}{\ensuremath{\reqs[\textsc{plan}]}\xspace}
\newcommand{\histreqs}{\ensuremath{\reqs[\textsc{hist}]}\xspace}
\newcommand{\conf}{\ensuremath{\alpha}\xspace}

\newcommand{\req}{r}
\newcommand{\rreq}{\tilde{r}}   

\newcommand{\oreq}{v(\req)}    
\newcommand{\areq}{\app(\req)}  
\newcommand{\treq}[1][\req]{t(#1)}    
\newcommand{\Treq}[1][\req]{T(#1)}    
\newcommand{\roreq}{v(\rreq)}    
\newcommand{\rareq}{\app(\rreq)}  
\newcommand{\treqs}{\reqs(t)}    

\NewDocumentCommand\greq{d<>}
    {\IfNoValueTF{#1}
        {\gapp[\areq]}
        {\gapp[#1]}}
\NewDocumentCommand\rgreq{d<>}
    {\IfNoValueTF{#1}
        {\gapp[\rareq]}
        {\gapp[#1]}}

\NewDocumentCommand\nreq{d<>O{i}}
    {\IfNoValueTF{#1}
        {\napp<\req>[#2]}
        {\napp<\req,#1>[#2]}}

\NewDocumentCommand\lreq{d<>O{ij}}
    {\IfNoValueTF{#1}
        {\lapp<\req>[#2]}
        {\lapp<\req,#1>[#2]}}

\NewDocumentCommand\dreq{sD<>{\req}O{t}}  
    {\IfBooleanTF{#1}
        {d(#2)}
        {d(#2, #3)}}
\NewDocumentCommand\rdreq{sD<>{\rreq}O{t}}  
    {\IfBooleanTF{#1}
        {d(#2)}
        {d(#2, #3)}}

\newcommand{\layertxt}{quantile\xspace}
\newcommand{\layertxts}{quantiles\xspace}
\newcommand{\layer}{p}
\newcommand{\Layer}{P}
\newcommand{\layers}{\Set P}

\NewDocumentCommand\Xreq{o}
    {\IfNoValueTF{#1}
        {\boldsymbol{x}}
        {\boldsymbol{\hat{x}}(#1)}}

\NewDocumentCommand\xreq{d<>O{\req}mm}
    {\IfNoValueTF{#1}
        {x^{#3}_{#4}(#2)}
        {x^{#3,#1}_{#4}(#2)}}

\NewDocumentCommand\ryreq{d<>O{\req}mm}
    {\IfNoValueTF{#1}
        {y^{#3}_{#4}(\tilde{#2})}
        {y^{#3,#1}_{#4}(\tilde{#2})}}

\NewDocumentCommand\xlayer{d<>O{\req}}
    {\IfNoValueTF{#1}
        {\underline{x}^{\layer}(\req)}
        {\underline{x}^{#1}(#2)}}

\NewDocumentCommand\Yreq{o}
    {\IfNoValueTF{#1}
        {\boldsymbol{y}}
        {\boldsymbol{y}(#1)}}

\NewDocumentCommand\yreq{d<>O{\rreq}mm}
    {\IfNoValueTF{#1}
        {y^{#3}_{#4}(#2)}
        {y^{#3,#1}_{#4}(#2)}}

\NewDocumentCommand\ylayer{d<>O{\rreq}}
    {\IfNoValueTF{#1}
        {\underline{y}^{\layer}(\rreq)}
        {\underline{y}^{#1}(#2)}}

\NewDocumentCommand\size{smO{t}o} 
    {\IfNoValueTF{#4}
        {\myop{load}}
        {\myop{load}^{#4}}
     \IfBooleanTF{#1}
        {(#2)}
        {\IfEq{#3}{t}
            {(#2, t)}  
            {(#2, #3)}}}

\newcommand{\res}[2][t, \Xreq]{\myop{Res}(#2, #1)}

\NewDocumentCommand\vnea{d()}
    {\ensuremath{\IfNoValueTF{#1}
        {\operatorname{VNEA}}
        {\operatorname{VNEA}(#1)}}}

\newcommand{\OPT}{\textsc{\textbf{s\scalebox{.8}{lot}o\scalebox{.8}{ff}}}\xspace}
\newcommand{\Greedy}{\textsc{\textbf{q\scalebox{.8}{uickG}}}\xspace}
\newcommand{\GreedyILP}{\textsc{\textbf{f\scalebox{.8}{ullG}}}\xspace}
\newcommand{\OLIVE}{\textsc{\textbf{o\scalebox{.8}{live}}}\xspace}

\usepackage{enumitem}
\usepackage{eso-pic}
\begin{document}
\AddToShipoutPictureBG*{%
  \AtTextLowerLeft{%
    \raisebox{-2\height}{%
      \parbox{\textwidth}{\centering%
      \copyright 2025 IEEE. Personal use of this material is permitted.\\ 
      Permission from IEEE must be obtained for all other uses, in any current or future media, including reprinting/republishing this material for advertising or promotional purposes, creating new collective works, for resale or redistribution to servers or lists, or reuse of any copyrighted component of this work in other works.} 
      \hspace{\columnsep}\makebox[\columnwidth]{}
    }%
  }%
}

\title{Plan-Based Scalable Online Virtual Network Embedding}


\author{
    \IEEEauthorblockN{Oleg Kolosov}
    \IEEEauthorblockA{Computer Science \\
    Technion\\
    Haifa, Israel 
    \href{mailto:kolosov@cs.technion.ac.il}{kolosov@campus.technion.ac.il} 
    }
\and
    \IEEEauthorblockN{David Breitgand}
    \IEEEauthorblockA{Hybrid Cloud \\ 
    IBM Research -- Haifa \\
    Haifa, Israel \\
    \href{mailto:davidbr@il.ibm.com}{davidbr@il.ibm.com}
    }
\and
    \IEEEauthorblockN{Dean H. Lorenz}
    \IEEEauthorblockA{Hybrid Cloud \\ 
    IBM Research -- Haifa \\
    Haifa, Israel \\
    \href{mailto:dean@il.ibm.com}{dean@il.ibm.com}
    }
\and
    \IEEEauthorblockN{Gala Yadgar}
    \IEEEauthorblockA{Computer Science \\
    Technion\\
    Haifa, Israel \\
    \href{mailto:gala@cs.technion.ac.il}{gala@cs.technion.ac.il}
    }
}
\author{
    \IEEEauthorblockN{%
    Oleg Kolosov\thanks{This work was partially funded by the IBM–Technion Research Collaboration, US-Israel BSF grant 2021613, and ISF grant 807/20.}\thanks{\textsuperscript{§}The authors are listed in alphabetical order except for the 1\textsuperscript{st} author.}\IEEEauthorrefmark{1},
    David Breitgand\IEEEauthorrefmark{2},
    Dean H. Lorenz\IEEEauthorrefmark{2},
    Gala Yadgar\IEEEauthorrefmark{1} 
    }
    \IEEEauthorblockA{
    \IEEEauthorrefmark{1}\textit{Technion}, Israel; 
    \IEEEauthorrefmark{2}\textit{IBM Research}, Israel\textsuperscript{§}
    }
}

\maketitle

\begin{abstract}

Network virtualization allows hosting applications with diverse computation and communication requirements on shared edge infrastructure. Given a set of requests for deploying virtualized applications, the edge provider has to deploy a maximum number of them to the underlying physical network, subject to capacity constraints. This challenge is known as the \textit{virtual network embedding (VNE)} problem: it models applications as virtual networks, where virtual nodes represent functions and virtual links represent communication between the virtual nodes.

All variants of VNE are known to be strongly NP-hard. Because of its centrality to network virtualization, VNE has been extensively studied. We focus on the online variant of VNE, in which deployment requests are not known in advance. This reflects the highly skewed and unpredictable demand intrinsic to the edge. Unfortunately, existing solutions to online VNE do not scale well with the number of requests per second and the physical topology size.

We propose a novel approach in which our new online algorithm, \OLIVE, leverages a nearly optimal embedding for an aggregated expected demand. This embedding is computed offline. It serves as a plan that \OLIVE uses as a guide for handling actual individual requests while dynamically compensating for deviations from the plan. We demonstrate that our solution can handle a number of requests per second greater by two orders of magnitude than the best results reported in the literature. Thus, it is particularly suitable for realistic edge environments.

\end{abstract}

\setlist[itemize]{leftmargin=*}
\setlist[enumerate]{leftmargin=*}
\def\sscale{.8}



\newcommand{\notationtable}{\begin{table}
    \caption{Notations}
    \label{tab:notations}
    \centering\setlength{\tabcolsep}{2pt}
    \footnotesize
    \begin{tabularx}{\columnwidth}{@{}lXr@{}}
        \toprule
        \textbf{Notation} & \textbf{Description} & \\ 
        \midrule
        $\gnet$ & substrate (physical network)
        & \multirow[b]{1}{*}{\llap{\color{blue}\shortstack[r]{Substrate}}} \\
        $\nnet \in \gnet, \lnet\in\gnet$ & node and link in $\gnet$ \\
        $\cost{\snet}$ & resource cost for element $s\in\gnet$ (node or link) \\
        $\capacity{\snet}$ & available resources on element $\snet\in\gnet$ (node/link) \\
        \midrule
        $a \in \allapps$ & an application
        & \multirow[b]{1}{*}{\llap{\color{blue}\shortstack[r]{Applications}}} \\
        $\gapp$ & virtual network topology of $a$ \\
        $\napp<a>, \lapp<a>, \user<a>$ & virtual node, virtual link, root node of $\gapp$ \\
        $\dapp{\sapp}$ & size of virtual element $\sapp\in\gapp$ (node or link) \\
        $\dnet{\sapp}{\snet}$ & (in)efficiency of serving $\sapp\in\gapp$ on $\snet\in\gnet$ \\
        \midrule
        $\req \in \reqs$ & an embedding request
        & \multirow[b]{1}{*}{\llap{\color{blue}\shortstack[r]{Requests}}} \\
        $\allt\kern-0.3em=\kern-0.2em\set{0, \Ldots, t, \Ldots, T\kern-0.3em-\kern-0.2em1}$ & overall time interval $\allt$, $t$ is a single time slot \\
        $\oreq, \areq$ & ingress location and application of request $\req$ \\
        $\dreq*$ & demand size of request $\req$ \\
        $\treq, \Treq \in \allt$ & arrival time, duration of request $\req$. $\req$ is \emph{active} for $\treq \le t < \treq + \Treq$\\
        $\treqs$ & active requests at time $t$ \\
        $\Psi(\req)$ & rejection cost (incurs if $\req$ is rejected) \\ 
        $\psi$ & rejection cost factor. $\Psi(\req) \triangleq \psi\dreq*\Treq$ \\
        \midrule
        $\Xreq(\reqs)$ & embedding of request set $\reqs$ 
        & \multirow[b]{1}{*}{\llap{\color{blue}\shortstack[r]{Embedding}}} \\
        $\xreq{\sapp}{\snet} \in \set{0,1}$ & $1$ iff $\Xreq$ maps $\sapp\in\gapp$ to $\snet\in\gnet$ for $\req\in\reqs$ \\
        $\layer \in \layers = \set{1, \Ldots, \Layer}$ & \layertxt of a request \\
        $\xlayer \in \set{0,1}$ & $1$ iff $\Xreq$ sets \layertxt of $\req\in\reqs$ to $\layer$ \\
        $\size{\snet}$ & load induced by $\Xreq$ on element $s\in\gnet$ at time $t$ \\
        $\Psi(\Xreq)$ & rejection cost (lost profit) for allocation $\Xreq$: $\sum_{\set{\req\in\reqs \mid \req \text{ is rejected by }\Xreq}} \Psi(\req) $ \\ 
        \bottomrule
    \end{tabularx}
    \vspace{-3mm}
\end{table}}

\section{Introduction}\label{sec:intro}
In edge computing, small-scale \dcs extend the cloud by bringing storage and compute resources closer to users. 
Application providers deploy their networked \textit{applications} (e.g., gaming and augmented reality) on edge \dcs, thus improving user quality of experience~\cite{etsi_mec}. 
Edge providers use \emph{network virtualization} to decouple deployment from the physical network and to isolate different applications from one another with respect to security and performance. This allows edge providers to utilize the same physical \textit{substrate} network for hosting applications with diverse topologies, communication, and computation requirements~\cite{LIM2023983}. 
Typical edge applications follow the micro-services design pattern~\cite{pranos,PASE}.
The micro-services are provided as \emph{virtual network functions (VNFs)}. A structured collection of VNFs forms an \emph{application}. The structure of an application is modeled as a \textit{virtual network (VN)} --- a graph whose nodes are the VNFs interconnected by \emph{virtual links} representing communication between the VNFs. 

\ignore{In practice, the edge operates under strict resource constraints and highly dynamic and unforeseen demand. \textit{Network virtualization} is crucial in the evolution of cloud and edge computing~\cite{etsi_mec}, enabling the same physical network to serve users with diverse communication and computation requirements. In this context, a \textit{virtual network (VN)} is a graph whose nodes are \textit{virtual network functions (VNFs)}, interconnected by \textit{virtual links}. VNFs and virtual links abstract hardware resource requirements from the underlying physical \textit{substrate network}. } 

In this model, the edge provider's goal is as follows. 
Given user requests to deploy applications, the edge provider needs to \textit{embed} their corresponding VNs onto the given physical substrate subject to computational and communication constraints. This fundamental problem is known in the literature as  
\textit{virtual network embedding (\vne)}. 

Each VN has physical capacity requirements for its VNFs and traffic demands for virtual links. Embedding a VN onto the substrate network entails finding a feasible mapping of each VNF to a substrate node and each virtual link to a \textit{path} in the substrate network. A feasible mapping must satisfy the demand requirements while adhering to the capacity constraints of the substrate nodes and links. Given a set of VN deployment requests, the \vne problem requires to determine which requests to accept, and how to allocate the substrate network resources to each accepted request. The goal is to minimize the rejection rate, or, conversely, maximize profit (i.e., gains obtained for successful embeddings~\cite{rost2019virtual}). This problem is known to be strongly NP-hard~\cite{Rexford2008}.

In \textit{online} \vne, VN requests arrive sequentially, whereas in \textit{offline} \vne, all VN requests are known in advance. Both offline~\cite{pranos,rost2019virtual,Rexford2008,ren2018optimal,tanto} and online~\cite{LIM2023983,nguyen2021distributed,rubio2024novel} problem variants have been extensively studied. Between the online and offline \vne, the online problem better reflects realistic scenarios, in which demand is unpredictable. However, existing online solutions face several significant limitations. 
Exact solutions based on \textit{integer linear programming (ILP)}  do not scale with the problem size (i.e., topology size and constraints) and request arrival rates~\cite{ExactVNE-Survey-2016,HeHZ-LCN2023,melo2013optimal,melo2015optimal,PASE}. Heuristic-based approaches~\cite{ChowdhuryRB-TON2012,Cheng-VNE2011,CaoZY-JCN-2019} offer better scalability in terms of problem complexity but have been evaluated under limited service rates, indicating that improving service rate scalability remains a challenge. Some methods attempt to enhance performance by imposing restrictions, such as delaying requests and serving them in batches.
AI-based solutions~\cite{LIM2023983,DVNE-DRL,DeepVine,wang2021proactive,gu2019deep,liu2023energy} outperform heuristic approaches in terms of service rate scalability but are highly sensitive to the quality of training data and model parameters. Furthermore, some AI-based solutions focus on optimizing specific aspects of the system (e.g., VNF embedding) while neglecting others, such as link embedding quality~\cite{DVNE-DRL,DeepVine}, potentially leading to sub-optimal solutions. Ultimately, the challenge lies in achieving both runtime scalability and optimality.

Many edge computing scenarios, e.g., 5G/6G, present extreme scalability challenges. For example, 5G requires scaling to the peak of one million connections per square kilometer~\cite{Liu-5GKPI-2020} with 20K simultaneously active sessions per square kilometer on average~\cite{OughtonFGB-Netherlands5G-2019}. Thus, developing scalable solutions to \vne is of great practical value.

In this paper, we propose a novel approach for the online \vne problem. The key idea is to use an offline \vne solution for expected aggregated demand (estimated from the observed history) as input to the online algorithm that uses this solution as a \textit{plan} for embedding. Our plan-based algorithm, \OLIVE (\textbf{O}n-\textbf{LI}ne \textbf{V}irtual network \textbf{E}mbedding), embeds actual requests as they arrive, while handling possible deviations from the expected demand. This approach supports service rates that are higher by two orders of magnitude than those reported for existing solutions.

In a typical network, there are relatively few applications but potentially many requests for each application~\cite{pranos}. 
Therefore, requests can be aggregated per substrate node to obtain an expected aggregated demand for each application at the that node. We use this aggregated demand as input to offline \vne. We solve it via \emph{linear program (LP) relaxation} to obtain a globally optimized embedding plan. Thanks to the LP formulation, even very large plans can be computed very quickly. Later, at the online phase, \OLIVE uses this LP relaxation to embed individual VN requests as they arrive and should be either feasibly mapped or rejected. Thus, the LP relaxation serves as an embedding plan helping to quickly find feasible embedding for the incoming request of a given class. 
A key unique property of our LP relaxation is that it guarantees a minimal allocation of the physical substrate capacity for each \textit{expected} request class. We show that this is property is essential when the solution is used as a plan.

Our plan-based algorithm, \OLIVE, handles momentary deviations from the expected demand, and therefore, from the optimal embedding plan. It dynamically reapportions physical capacity to ensure that no VN request class is starved as long as unused capacity is available from other underutilized classes. To achieve this, \OLIVE ``borrows’’ capacity and redistributes it among the classes. If a previously underutilized class becomes over-utilized, any ``borrowed'' capacity above the guaranteed share of another class is preempted. 

We show that our approach produces near-optimal results and significantly outperforms an online strategy that does not use planning. Local optimization methods, such as greedy heuristics commonly used in practice, might get stuck in local minima. Our solution avoids these pitfalls by virtue of the globally optimized plan.
This combination of efficiently computed offline planning with dynamic corrections during online embedding is the main novelty of \OLIVE. 
Our specific contributions are as follows.

\begin{itemize}
    \item We present \OLIVE\footnote{The source cost is available at \url{https://github.com/olekol33/VNE}}, a novel nearly-optimal plan-based heuristic to solve online \vne that provides guaranteed allocation for expected demand (\cref{sec:solution}).
    \item Using a wide range of synthetic and real topologies and workloads, we show that
    \OLIVE scales linearly with the problem size and has short running times, making it a practical solution. Specifically, simulations show that \OLIVE supports a request rate per second that is \textit{two orders of magnitude higher} than those reported in the literature (\cref{sec:eval}). 


\end{itemize}

\begin{figure*}[h!]
    \def\scale{0.3}
    \subfloat[Applications, requests]{%
    \footnotesize\centering\setlength{\tabcolsep}{1pt}
    \begin{tabular}[b]{lccccc}
    \multicolumn{6}{@{}c@{}}{
    \includegraphics[page=7,scale=\scale,trim=0.5cm 13.9cm 26cm 0.5cm,clip]{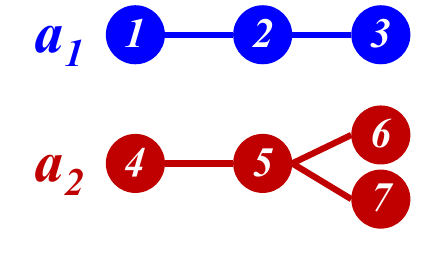}} \\
    &$\treq$ & $\Treq$ & $\oreq$ & $\areq$ & $\dreq*$ \\
    \midrule
    $r_1$ & 1 & 4 & $A$ & $a_1$ & $1$ \\
    $r_2$ & 2 & 2 & $B$ & $a_2$ & $1$ \\
    $r_3$ & 3 & 3 & $D$ & $a_2$ & $2$ \\
    \end{tabular}  
    \label{fig:onvne6}}
    \hfill
    \subfloat[Substrate $\gnet$]{\includegraphics[page=6,scale=\scale,trim=0.5cm 9.9cm 26cm 0.5cm,clip]{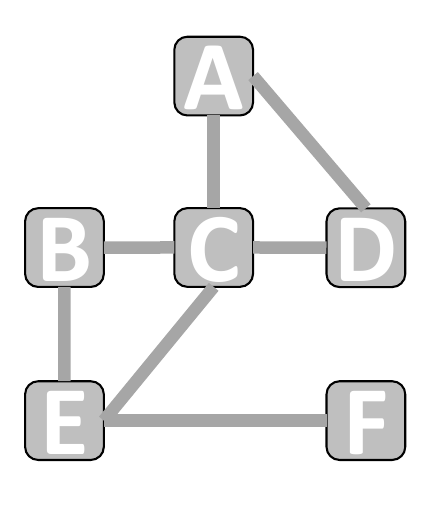}}\label{fig:onvne1}
    \hfill
    \subfloat[$t=1$]{\includegraphics[page=8,scale=\scale,trim=0.5cm 9.9cm 26cm 0.5cm,clip]{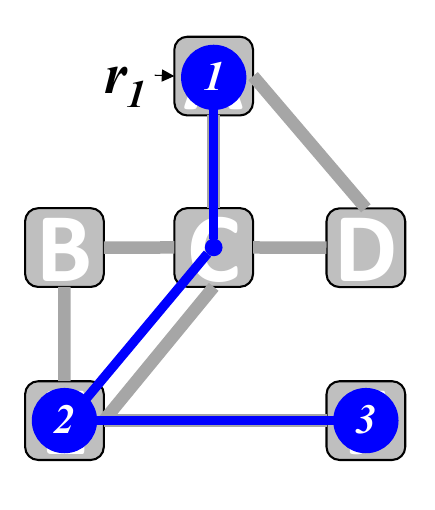}}\label{fig:onvne2}
    \hfill
    \subfloat[$t=2$]{\includegraphics[page=9,scale=\scale,trim=0.5cm 9.9cm 26cm 0.5cm,clip]{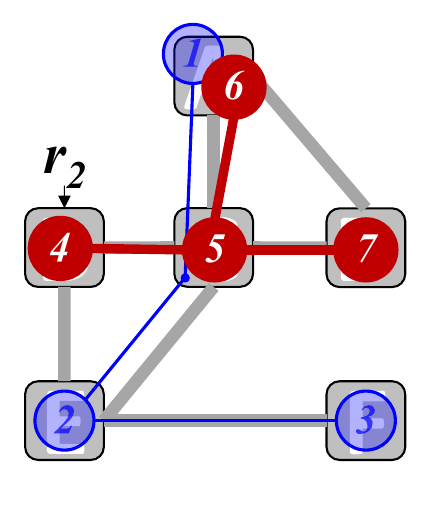}}\label{fig:onvne3}
    \hfill
    \subfloat[$t=3$]{\includegraphics[page=10,scale=\scale,trim=0.5cm 9.9cm 26cm 0.5cm,clip]{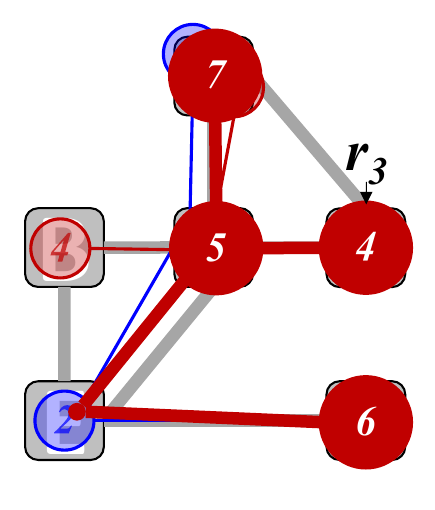}}\label{fig:onvne4}
    \hfill
    \subfloat[$t=4$]{\includegraphics[page=11,scale=\scale,trim=0.5cm 9.9cm 26cm 0.5cm,clip]{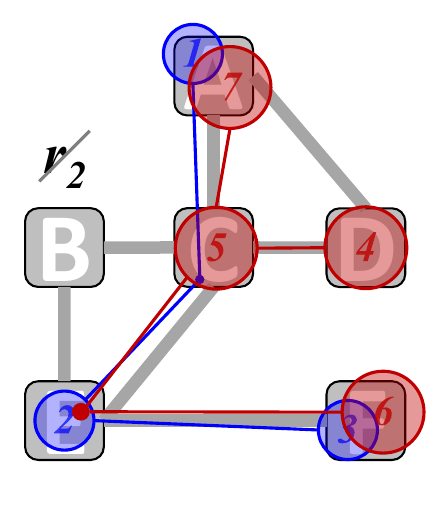}}\label{fig:onvne5}
    
    \caption{Online embedding example. Possible applications are $\allapps = \set{a_1, a_2}$. Requests $r_2$, $r_3$ are both for application $a_2$, but with different ingress and size. $r_1$,$r_2$, and $r_3$ are embedded at time slots 1,2, and 3, respectively. At $t=4$ $r_2$ departs.
    }
    \label{fig:onembed}
\end{figure*}

\notationtable

\section{Online Virtual Network Embedding Problem}\label{sec:problem}
In this section, we introduce definitions and notations (\cref{tab:notations}), and present the \Onvne problem and its relation to the classical offline problem.

\subsection{Definitions and Notations}\label{sec:preliminaries}

\T{Physical substrate network} The substrate network is a set of interconnected \dcs. It is modeled as a graph $\gnet$, where each node $v \in \gnet$ represents a \dc and each link $\lnet\in\gnet$ represents a connection between \dcs. The functions $\cost{\snet}$ and $\capacity{\snet}$ define the usage cost (e.g., energy consumption or monetary expenses) and capacity for each element $\snet \in \gnet$ (node or link). As VN requests arrive and get embedded, the overall embedding on each element $\snet \in \gnet$ induces a \emph{load}, representing the element's available capacity used by the embeddings. 

\T{Applications} Each application $\app \in \allapps$ is defined by its virtual network topology graph, $\gapp$. Each node $\napp \in \gapp[a]$ represents a VNF $i$ in $a$; a link $\lapp$ represents interaction between VNFs $i,j$. Each element  $\sapp\in\gapp$ (either node or link) has size $\dapp{\sapp}$, which represents this element's resource requirements.

Each topology $a \in \allapps$ has a special node $\user<a>$ that represents a user. Virtual link(s) between $\user<a>$ and other nodes 
represent the direct interaction between the user and the functions represented by these nodes. In this paper, we consider tree and chain network topologies with $\user<a>$ as their \emph{root} nodes. Since $q=\user<a>$ only represents $a$'s ingress point, we set $\dapp{\user} \gets 0$. 

\T{Time} We consider a finite overall time interval $\allt$, consisting of discrete time slots $t \in \allt$.

\T{Requests} Each online request $\req \in \reqs$ for embedding an application $\areq \in \allapps$ is generated at time $\treq$ by a user located at substrate node $\oreq$. A request's demand size is denoted by $\dreq* \in \Rplus$. Each request is \emph{active} for a duration $\Treq$ between its \emph{arrival} time $\treq$ and its \emph{departure} time $\treq+\Treq$. The set of active requests at time $t$ is denoted by $\treqs$. The parameters $\areq, \oreq, \dreq*$ for each request are known only upon its arrival time and the duration $\Treq$ is known only upon its departure. 

\T{Embedding} A request is \emph{embedded} if all virtual nodes in $a$ are mapped on physical nodes in $\gnet$ and all virtual links are mapped on paths in $\gnet$. Otherwise, the request is \emph{rejected}. 
\T{Validity} The allocation for $\req$ is defined by $\Xreq(\req)$, where $\xreq{\sapp}{\snet} = 1$ iff $\Xreq(\req)$ maps virtual element $\sapp\in\gapp$ to $\snet\in\gnet$ and $0$ otherwise. A \textit{valid} embedding is \emph{unsplittable}, thus $\xreq{\sapp}{\snet} = 1$ for exactly one substrate element $\snet$ 
(for rejected requests it is 0 for all $\snet$). The allocation $\Xreq(\req)$ is set at $\treq$, after that requests cannot be \emph{re-}allocated to new substrate resources.

\T{Resource Consumption} An allocated request $\req$ consumes substrate resources while it is active. For an allocation $\Xreq(\req)$, the amount of resources consumed by virtual element $\sapp$ on substrate element $\snet$ is defined as: 
\begin{equation}\label{eq:loadsq}
\size*{\Xreq(\req), \sapp, \snet} \triangleq \xreq{\sapp}{\snet} \cdot \dreq* \cdot \dapp{\sapp} \cdot \dnet{\sapp}{\snet},  
\end{equation}
where $\dnet{\sapp}{\snet}$ is the \emph{(in)efficiency} coefficient for allocating $\sapp$ on $\snet$. The combination of $\dapp{\sapp}$ and $\dnet{\sapp}{\snet}$ allows flexibility in modeling applications and factoring in placement policies. A higher $\dnet{\sapp}{\snet}$ value means $\sapp$ takes up more resources on $\snet$. A low value suggests that it is preferable to place $\sapp$ on $\snet$; for example, placing a packet-processing function on a node with hardware acceleration support. Extremely high $\dnet{\sapp}{\snet}$ values can be used to prevent mapping $\sapp$ to $\snet$, for reasons such as privacy, compliance, or performance constraints. 

\begin{figure}
\begin{problem}{\COnvne}{\onvne}\label[onvne]{prob:onvne}%
\textbf{Given} $\reqs, \allapps, \gnet$, optimization~criteria
\par\textbf{Sort} $\reqs$ by request arrival times. 
\begin{quote}
Distinct requests have distinct arrival times ($i \neq j$ implies $\treq[\req_i] \neq \treq[\req_j]$); equal arrival times are ordered arbitrarily.
\end{quote}
\par\textbf{Find} $\Xreq$ by processing the requests sequentially.    
For each $\req \in \reqs$:
\begin{enumerate}
    \item Given  $r$ and previous allocation $\set{\Xreq(\req') \text{ s.t. } \treq[\req'] < \treq }$
    \item Find an allocation $\Xreq(\req)$ (or \emph{reject} $\req$, by setting $\Xreq(\req) \gets 0$)
\end{enumerate}

\par\textbf{Such that}
\begin{enumerate}
    \item $\Xreq$ is \emph{feasible} for each $t \in \allt$
    \item $\Xreq$ is \emph{optimal} w.r.t. the given optimization criteria
\end{enumerate}
\end{problem}
\vspace{-\baselineskip}
\caption{Online \vne}\label{fig:onvne}
\vspace{-.5\baselineskip}
\end{figure}

\subsection{\COnvne}\label{sec:onvne}

\Cref{fig:onembed} depicts an example of online embedding of three requests. Requests are processed in their arrival order and each request is embedded onto the substrate network. The resulting allocation, over all time slots, must be feasible and optimal.
We formally define \onvne in \cref{fig:onvne}.

\T{Feasibility} An allocation is feasible if resource consumption does not exceed the substrate capacity constraints for all time slots. That is,
$\size{\Xreq, \snet} \leq \capacity{\snet} \quad \forall \snet \in \gnet, \forall t \in \allt$,
where $\size{\Xreq, \snet}$ is the aggregated form of \cref{eq:loadsq}:
\begin{equation}\label{eq:loads}
    \size{\Xreq, \snet} \triangleq 
    \smashoperator{\sum_{\req \in \treqs}} 
    \dreq* \!\smashoperator{\sum_{\sapp \in \greq}} 
    \xreq{\sapp}{\snet} \dapp{\sapp} \dnet{\sapp}{\snet}
\end{equation}
\ignore{Note that the (in)efficiency  coefficients $\dnet{\sapp}{\snet}$ allow feasibility to include placement restrictions (equivalent to ``valid mapping'' in \cite{RostSchmidt-ToN2020})}

\T{Optimality} The objectives typically studied for \vne are to minimize resource allocations or to maximize the profit. The former minimizes the cost of consumed resources 
\begin{equation}\label{eq:costres}
\cost[\gnet]{\Xreq} \triangleq \sum_{t \in \allt}\sum_{\snet \in \gnet} \size{\Xreq, \snet} \cost{\snet}.
\end{equation}
The latter objective assigns a \emph{benefit} for each allocated request and attempts to maximize the overall benefit. We convert profit maximization to cost minimization by assigning a rejection \emph{penalty factor} $\psi(\req)$ for each request. The overall rejection cost (lost profit) is:
\begin{equation}\label{eq:costreject}
\Psi(\Xreq) = \cost[\textit{rejections}]{\Xreq} \triangleq \smashoperator{\sum_{\set{\req \in \reqs \mid \Xreq(\req) = 0}}} \dreq* \Treq \psi(\req).
\end{equation}
In the following we use minimization of 
$\cost[\gnet]{\Xreq} + \Psi(\Xreq)$
as the optimization criteria.

\subsection{\COfvne}\label{sec:ofvne}

The traditional offline version of \vne (\ofvne) is very similar to \onvne, with $\allt = \set{0}$. There is only one time slot; all requests arrive at time $t=0$, and the goal is to find the optimal feasible embedding, as defined for \onvne. 

The main difference is that in \ofvne there is no complete order on request arrival times and there is no need to process requests in sequential order. That is, all requests are known simultaneously at time $t=0$ and that knowledge can be utilized to find an optimal $\Xreq$.

\newcommand{\alg}
{\begin{algorithm}
\caption{\textsc{Olive}}\label{alg:olive}\label[olive]{olive}
\Small
\begin{algorithmic}[1]
    \Require{$\gnet, \allapps, \reqs, \Yreq$}
    \Ensure{An embedding $\Xreq$ that is a feasible solution to \cref{prob:onvne}}
    \Statex \vspace{0.3\baselineskip}\hrule
    \State $\donereqs \gets \emptyset$, $\planreqs \gets \emptyset$
    \State Initialize $\res[0,\emptyset]{\gnet}$ using \eref{eq:ress} \Comment{residual capacity of substrate $\gnet$}
    \State Initialize $\res[0,\emptyset]{\Yreq}$ using \eref{eq:resy} \Comment{residual capacity of \PLAN $\Yreq$}
    \For{$t \in \allt$}
        \State Update $\res{\gnet}, \res{\Yreq}$ \Comment{handle departing requests}
        \For{$\req \in \reqs$, s.t. $\treq = t$} \Comment{process one by one in arrival order} \label{lin:newreq}
            \State $\Xreq(\req), \textit{planned} \gets $ \label{lin:eight} 
            \Call{PlanEmbed}{$\Yreq, \req$}
            \If{\textit{planned} and $\Xreq(\req) > \res{\gnet}$}\Comment{no available capacity}\label{lin:premept}
                \State\Call{Preempt}{$\req, \donereqs, \planreqs, \Xreq(\req)$} 
            \EndIf
            \If{$\Xreq(\req) = \emptyset$} \label{lin:nox}
                \State $\Xreq(\req) \gets $ \Call{GreedyEmbed}{$\req$}

            \EndIf
            \If{$\Xreq(\req) \neq \emptyset$}\Comment{Implies $\Xreq(\req) \le \res{\gnet}$}
                \State \Call{Allocate}{$\planreqs, \Xreq(\req)$, \textit{planned}}
                \Comment{\emph{accept $\req$}}
            \Else \State$\xreq{\user}{\oreq} \gets 0$ \Comment{\emph{reject $\req$}}
            \EndIf
        \EndFor
        \State $\donereqs \gets \donereqs \cup \set{\req}$
    \EndFor
    \State \Return $\Xreq$
    \Statex \vspace{0.3\baselineskip}\hrule 
    \Function{Allocate}{$\planreqs, \Xreq(\req), \textit{planned}\in\set{\textit{True}, \textit{False}}$}
        \State $\xreq{\user}{\oreq} \gets 1$ \Comment{\emph{allocate $\req$}}
        \State $\Xreq \gets \Xreq \cup \set{\Xreq(\req)}$, update $\res{\gnet}$ \label{lin:twentyeight}
        \If{\textit{planned=True}}
            \State $\planreqs \gets \planreqs \cup \set{\req}$, update $\res{\Yreq}$ \label{lin:thirty}  
        \EndIf            
    \EndFunction
    \Statex \vspace{0.3\baselineskip}\hrule 
    \Function{PlanEmbed}{$\Yreq, \req$}
        \State Find $\rreq_{\areq, \oreq}$, s.t., $\req \in \rreq$ \Comment{Find aggregate request set of $\req$}
        \If{$\exists \Xreq[\req] \text{, s.t. } \Xreq[\req] \le \res{\Yreq[\rreq]}$}\Comment{fits in residual plan}
            \State\Return $\Xreq[\req]$, \textit{planned=True}\label{lin:plannedok}
        \EndIf
        \If{$\exists \Xreq[\req]$, $0 < \alpha < 1$, such that\label{lin:thirtyeight}\Comment{partial fit}
        \\\hspace*{\algorithmicindent}\phantom{\textbf{if}} 
        $\alpha \cdot \Xreq[\req] \le\res{\Yreq[\rreq]}$ 
        \textbf{and} 
        $\Xreq[\req] \le \res{\gnet}$
        \\\hspace*{\algorithmicindent}\phantom{\textbf{if}}}
            \Return $\Xreq[\req], \textit{planned=False}$\label{lin:notplannedok}
        \EndIf
        \State \Return $\emptyset, \textit{planned=False}$        
    \EndFunction
    \Statex \vspace{0.3\baselineskip}\hrule 
    \Function{GreedyEmbed}{$\req$}
        \State $\Xreq^{\textsc{feasible}} \triangleq$ all $\Xreq[\req]$, s.t. $ \Xreq[\req] \le \res{\gnet}$ 
        \State $\Xreq^{\textsc{collocated}} \triangleq$ all $\Xreq[\req]$, s.t. $\xreq{i}{\snet} = \xreq{j}{\snet} \quad \forall i,j \neq \user \in \gapp$
        \State \Return $\Xreq[\req] \in \Xreq^{\textsc{feasible}} \cap \Xreq^{\textsc{collocated}}$ with minimal cost
    \EndFunction
    \Statex \vspace{0.3\baselineskip}\hrule 
    \Function{Preempt}{$\req, \donereqs, \planreqs, \Xreq(\req)$}
        \State \Shortstack{Find $\reqs' \subseteq \donereqs \setminus \planreqs$, s.t., 
        freeing resources $\set{\Xreq[\req']}_{\req' \in \reqs'}$\\ 
        would allow allocation of $\Xreq(\req)$, i.e., $\Xreq(\req) < \res{\gnet}$}
        \State $\xreq[\req']{\user}{v(\req')} \gets 0 \quad \forall \req' \in \reqs'$ \Comment{\emph{reject $\req'$}}
        \State Update $\res{\gnet}$
    \EndFunction
\end{algorithmic}
\vspace{-1mm}
\end{algorithm}
}

\newcommand{\planalg}
{\begin{algorithm}
\caption{Solve \COnvne}\label{alg:plan}\label[plan]{plan}
\Small
\begin{algorithmic}[1]
    \Require{$\gnet, \allapps, \reqs, \histreqs, \conf$}
    \Ensure{A feasible embedding $\Xreq$ that is a solution to \cref{prob:onvne}}
    \State Construct an offline \vne problem instance, $\ofvne(\rreqs, \allapps, \gnet)$, where $\rreqs \gets$ time-aggregation of $\histreqs$, using \cref{eq:agg,eq:bootstrap}     \label{lin:agg}\Comment{\cref{sec:timeagg}}
    \State $\Yreq[\rreqs] \gets \text{ Solve }\cref{prob:pvne}(\gnet, \allapps, \rreqs)$ \label{lin:retplan} \Comment{\cref{sec:planvne}}
    \State \Return \Call{\cref{olive}}{$\gnet, \allapps, \reqs, \Yreq[\rreqs]$}\Comment{\cref{sec:olive}}
\end{algorithmic}
\end{algorithm}%
}

\section{Our Solution}\label{sec:solution}

\begin{figure}[h]
    \centering
    \vspace{-3mm}
    \includegraphics[page=4,scale=0.3,trim=4.5cm 2.9cm 4.1cm 3.9cm,clip]{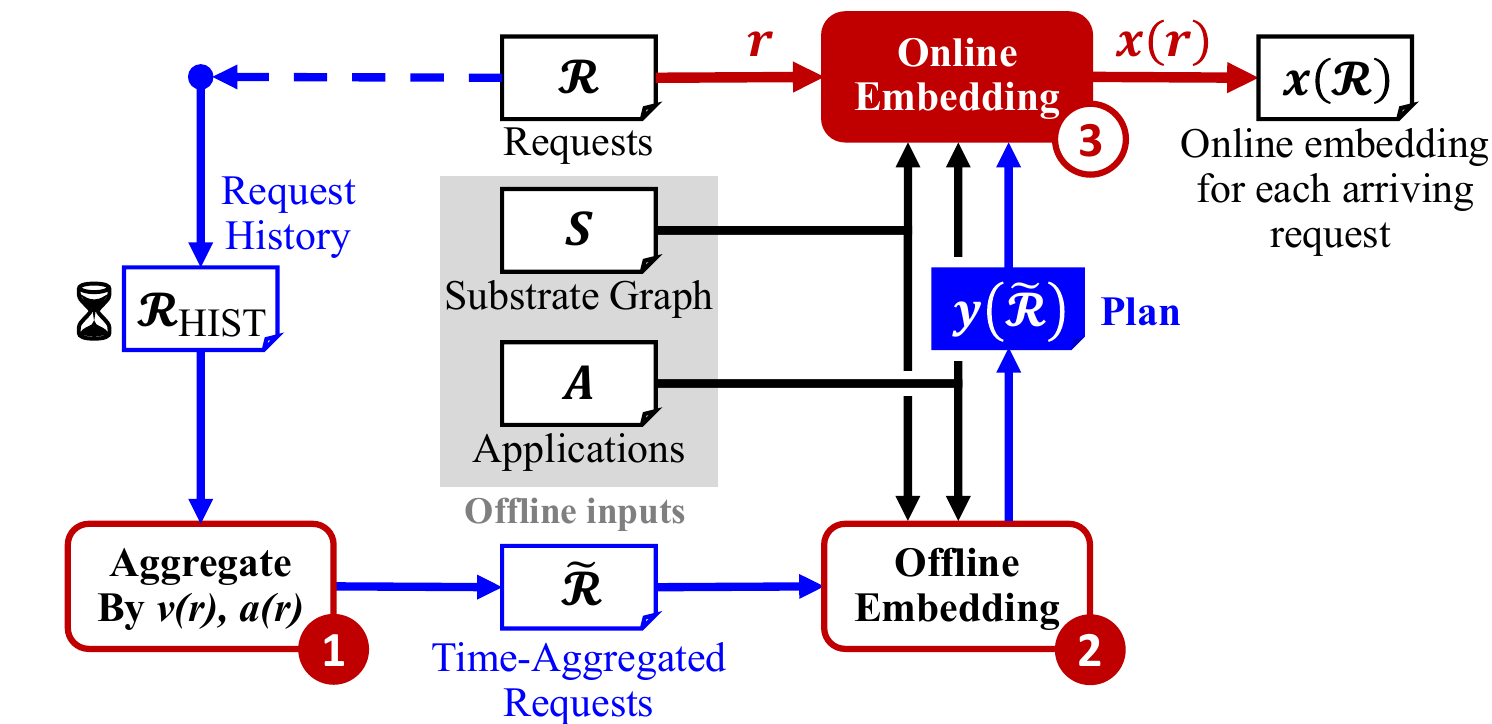}
    \caption{Solution overview}
    \label{fig:overview}
    \vspace{-3mm}
\end{figure}

In this section, we present our approach to the online \vne problem. \Cref{fig:overview} and~\cref{alg:plan} summarize the three steps comprising our solution:
\begin{enumerate}
    \item Use a history of requests, $\histreqs$, to create a \emph{time-aggregated} request set $\rreqs$ and an \ofvne instance (\cref{sec:timeagg}). Inspired by~\cite{pranos}, in our \ofvne construction, we classify requests by application and user location and create an aggregate demand for each class.
   \item Solve a novel relaxed \ofvne for $\rreqs$, called \cref{prob:pvne}, to produce a \emph{plan} $\Yreq$ (\cref{sec:planvne}). A novel feature introduced in \cref{prob:pvne} prevents an excessive rejection rate for any class of requests.
    \item Use $\Yreq$ to solve $\onvne$, allocating requests as they arrive using \OLIVE (\cref{sec:olive}). This usage of the offline plan for solving the online problem is the key innovation of this work. As our evaluation study confirms, utilizing the plan improves both the speed and the quality of request embedding.
\end{enumerate}

\planalg

\ignore{
First, we introduce a new aggregation method that takes into account the distribution of demand over time (\cref{sec:timeagg}). 
 Then we provide a new LP formulation,  (\cref{sec:planvne}) for \ofvne that adds decisions variables to extend the coverage of the produced plan over all request classes.} 
\ignore{
Lastly, we develop an entirely new algorithm, \cref{olive}, to utilize the plan for online embedding of requests (\cref{sec:olive}), including handling of deviations from the plan. 
\Cref{alg:plan} summarizes our approach. \OK{TODO: rewrite}}

\subsection{Time-aggregation}\label{sec:timeagg}

As a first step to compute an embedding plan, \cref{alg:plan} constructs an instance of the \ofvne problem (\cref{sec:ofvne}). We obtain the input request set for this instance by the following procedure. First, we aggregate a history of requests, $\histreqs$, both by application and user location and over time to represent a \textit{maximal} peak demand for each request class (as if all requests of a given class arrive simultaneously). 
A specific methodology for the selection of requests to be included into $\histreqs$ is outside this paper's scope. \Cref{alg:plan} assumes that $\histreqs$ is a random sample coming from an unknown underlying distribution corresponding to a stationary process. Thus, it expects the online demand to be ``drawn'' from the same distribution.\footnote{ In~\cref{sec:eval}, we also explore the behavior of \OLIVE without this assumption.} We say that the \textit{online} demand conforms to the expectations from the history $\histreqs$ if the observed percentile $P_{\conf}$ of the online demand falls within the $95\%$ confidence interval of the estimated (i.e., expected) percentile $\hat{P}_{\conf}$, computed for the empirical cumulative distribution function (ECDF) of $\histreqs$. Since the percentile of a sample, $P_{\conf}$, is a random variable itself, we estimate it from $\histreqs$ by the well-known and widely used statistical technique of \emph{bootstrapping}~\cite{bootstrap}. In our evaluation, we use the value of the $\hat{P}_{80}$ estimated from the request history $\histreqs$ as anticipated peak aggregated demand. The rationale for using $\hat{P}_{80}$ rather than the full peak demand, $\hat{P}_{100}$, is to avoid over-provisioning.

\ignore{
This set shares statistical properties with the online requests, e.g., a set of requests immediately preceding the online requests or those collected under similar conditions (e.g., same time of day). \DB{I suggest removing the previous sentence.} \OK{OK, add footnote to prepare reader for evaluation with/without statistical properties. Verify text below doesn't heavily rely on this removed assumption. Actual (online) requests are expected demand relative to history of requests.}}
  
More formally, we group $\histreqs$ into an aggregate request set $\rreqs = \set{\rreq_{a,v}}_{\app \in \allapps, v \in \gnet}$ where 
\begin{equation}
\begin{aligned}
\rreq_{a,v} &= \set{r \in \histreqs \mid \areq=a \And \oreq=v} 
\end{aligned}\label{eq:agg}
\end{equation}
We set $\app(\rreq_{a,v}) = a$ and $v(\rreq_{a,v}) = v$. 
Since all requests $\req \in \rreq_{a,v}$, share the same application $a$ and the same user location $v$, they share the same placement constraints, virtual element sizes, and (in)efficiency coefficients. 

The total demand of the requests in $\rreq$ at time $t$ is given by 
$$
\rdreq \triangleq \smashoperator{\sum_{\req \in \rreq \cap \treqs}}\dreq*
$$
We define the time-independent \textit{expected} demand $\rdreq*$ as the estimated $\conf$ percentile of $\histreqs$ for $\rdreq$
\begin{equation}
\rdreq* \gets \hat{P}_{\conf}( \set{\rdreq}_{t})
\quad\quad\blue{\substack{\forall \rreq \in \rreqs}}
\label{eq:bootstrap}
\end{equation}
\ignore{
We apply \emph{bootstrap}~\cite{bootstrap} to estimate a desired target percentile $\conf$ for the aggregated demand. For instance, for $\conf=80\%$, we seek an estimated $\rdreq*$ that covers $\rdreq$ for 80\% of the time slots. $\rdreq*$ is estimated by drawing with replacement from the population of time-dependent demands $\rdreq$ and taking the $\conf$ percentile. This resampling is repeated enough times to find an estimate within a 95\% confidence interval }

While the embedding of each $\req \in \rreq$ may not be split, the aggregate demand 
$\rdreq*$ is \emph{splittable}. The aggregate request $\rreq$ represents multiple individual requests which can each have a different allocation. Moreover, the allocation for each $\rreq$ is only used as a guideline for the online embedding process.
It should be noted that each individual request is small compared to the size of the aggregated request.
Thus, we may utilize a relaxed solution when solving for $\rreqs$.

\subsection{Plan creation by solving \textsc{off-vne}}\label{sec:planvne}
\cref{prob:pvne} (\cref{fig:pvne}) presents our novel \ofvne LP formulation used to obtain an embedding plan. The optimization goal is to minimize overall cost, as defined in \cref{eq:ycost}, comprising resource costs and rejection costs. For each $\rreq$, the continuous decision variable $\yreq{\sapp}{\snet}$, defined in \cref{eq:ymilp}, gives the fraction of $\rdreq*$ for which $\sapp$ is mapped onto $\snet$. Thus, solving \cref{prob:pvne} produces a \textit{splittable} plan to solve \onvne. 

If no feasible solution exists to embed all aggregated requests, the solver may produce a solution that fully allocates some aggregated requests while completely rejecting others. Such unbalanced embeddings are not desirable as a plan for \onvne, as they provide no embedding guidance for online requests that belong to an unallocated aggregate demand. We address this using \emph{rejection \layertxts} (not to be confused with the percentiles of the expected aggregated demand discussed in the previous section), a novel method to prevent starvation of request classes. The approach assigns progressively increasing rejection costs, based on the fraction of rejected traffic for each request class. Each request's aggregated demand is divided into $\Layer$ equal parts, where each fraction of $1/\Layer$ of $\rdreq*$ is associated to a different \layertxt. The rejection penalty factor for the rejected traffic of \layertxt $\layer$ is $\psi(\rreq, \layer) \triangleq \layer \cdot \psi(\rreq)$. The \layertxt association is added as decision variables to the LP formulation (used only for handling rejected traffic).  

The assignment of \layertxts can be viewed as a ``water-filling'' method for equalizing rejection rates among all aggregate requests. When two requests compete for the same resources and some demand must be rejected, the cost optimization will prioritize rejecting demand with a lower associated rejection cost, namely with a lower \layertxt. Thus, requests that already have a substantial portion of their demand rejected will incur higher costs for additional rejections, as they must use higher \layertxts.

The load on each element $\snet\in\gnet$ is given by \cref{eq:ytsize}; an offline, time-independent version of \cref{eq:loads}. 
The feasibility constraints (element capacity) are given in \cref{eq:ycapacity}.
\Cref{eq:ymilpalt} defines \layertxts; 
$\ylayer \le 1/\Layer$ is the fraction of $\rdreq*$ that is rejected and assigned to \layertxt $\layer$. 
Rejection cost, defined in \cref{eq:ytreject}, is incurred for $\rreq$ only for its rejected traffic, as indicated by non-zero \layertxt assignments. The rejection cost is higher for the fraction of demand that is assigned a higher \layertxt value. We use a fixed base rejection factor $\psi(\rreq) = \psi$.

The allocated fraction of $\rreq$, computed by \cref{eq:yvalidroot} as the complement of the rejected traffic, is assigned to the root $\user<a>$ of its application $\rareq$ at the ingress point of the request $\roreq$. As stated in \cref{eq:yvalidroot1}, $\user$ must not be assigned on any other substrate node. 
Finally, flow preservation is defined in \cref{eq:yvalidpath}.

\begin{figure}
\begin{problem}{\CPvne}{\pvne}%
\label[pvne]{prob:pvne}
\vspace{-1.8\baselineskip}
\begin{align}
\shortintertext{\blue{\textbf{Given}} $\gnet, \allapps, \rreqs$,~\blue{\textbf{minimize:}}}
\label{eq:ycost}
\cost{\Yreq} &\triangleq \smashoperator{\sum_{\mathclap{\snet \in \gnet}}} \size*{\snet} \cost{\snet} + \Psi
& \phantom{MMM}{}%
\shortintertext{\blue{\textbf{where:}}}
\label{eq:ytsize} 
\size*{\snet} &\triangleq \sum_{\rreq \in \rreqs} \rdreq* \smashoperator{\sum_{\sapp \in \rgreq}} \yreq{\sapp}{\snet} \dapp{\sapp} \dnet{\sapp}{\snet}
& \mathllap{\blue{\substack{\hfill\forall \snet \in \gnet}}} 
\\\label{eq:ytreject} 
\Psi &\triangleq  \psi \smashoperator{\sum_{\req \in \rreqs}} \rdreq* \smashoperator{\sum_{\layer \in \layers}}\layer \cdot\ylayer 
& \mathllap{\blue{\substack{\hfill\layers = \set{1, \ldots, \Layer}}}}
\shortintertext{\blue{\textbf{such that}} $\Yreq$ \blue{\textbf{satisfies} $\forall \req \in \rreqs$:}}
\label{eq:ymilp} 
\yreq{\sapp}{\snet} &\in \range{0,1} 
&\mathllap{\blue{\substack{\hfill\forall \snet \in \gnet\\\hfill\forall \sapp \in \rgreq}}} 
\\\label{eq:yvalidroot1}
\yreq{\user}{v} &= 0 &\mathllap{\blue{\substack{\forall v\neq\roreq}}}
\\\label{eq:ymilpalt} 
\ylayer &\in \range{0,1 / \Layer}
&\mathllap{\blue{\substack{\hfill\forall \layer \in \layers\\\layers = \set{1, \ldots, \Layer}}}} 
\\\label{eq:yvalidroot} 
\yreq{\user}{\roreq} &= 1 - \smashoperator{\sum_{\layer \in \layers}}\ylayer
&\mathllap{\blue{\substack{
\layers = \set{1, \ldots, \Layer}}}}
\\\label{eq:yvalidpath} 
\yreq{j}{v} &= \yreq{i}{v} + \sum_{\mathclap{\lnet[uv] \in \gnet}} \yreq{ij}{uv} - \sum_{\mathclap{\lnet \in \gnet}} \yreq{ij}{vw}
&\mathllap{\blue{\substack{\hfill\forall v \in \gnet\\\hfill\forall \lapp \in \greq}}} 
\\\label{eq:ycapacity} 
\capacity{\snet} &\ge \size*{\snet}
& \mathllap{\blue{\substack{\hfill\forall \snet \in \gnet}}} 
\end{align}
\end{problem}
\vspace{-1\baselineskip}
\caption{LP formulation for \pvne}\label{fig:pvne}
\vspace{-3mm}
\end{figure}

\subsection{Online embedding utilizing the plan}\label{sec:olive}
\alg
\Cref{alg:olive} presents \OLIVE, our algorithm for online processing of incoming requests. \OLIVE uses a plan $\Yreq$ as a guideline for allocating online requests, ensuring balanced allocation across all requests and adhering to placement constraints. Whenever actual demand deviates from the expected demand $\rdreq*$ for which the embedding plan is calculated, \OLIVE employs ad hoc compensatory mechanisms to minimize rejection rate and cost, as will be described shortly. 

As online requests are being served sequentially in their order of arrival, the load $\size{\snet}$ on every substrate element changes dynamically. 
\OLIVE tracks the \textit{residual} capacities on the substrate elements $\snet \in \gnet$ to ensure that embedding of the new requests does not violate feasibility constraints. \ignore{A request element $\sapp\in\gapp$ is allocated on $\snet$ iff sufficient residual capacity is available on $\snet$ to support $\sapp$'s demand.}

\donereqs denotes the already processed requests. The substrate's residual capacity at time $t$ is defined in \eref{eq:ress}. \ignore{It is similar to the calculation in \eref{eq:loads}, but takes into account only processed requests, i.e., only \donereqs. }
\begin{equation}\label{eq:ress}
    \res{\gnet} \triangleq 
    \Bigl\{ \capacity{\snet} - \!\!\smashoperator{\sum_{\req \in \donereqs \cap \treqs}} 
    \dreq* \sum_{\sapp \in \mathrlap{\greq}} 
    \xreq{\sapp}{\snet} \dapp{\sapp} \dnet{\sapp}{\snet}
    \Bigr\}_{\snet\in\gnet}
\end{equation}
We define a \textit{residual plan} for a plan $\Yreq$ in \eref{eq:resy}.  
Given the requests that were already allocated according to $\Yreq$ (denoted by \planreqs), $\res{\Yreq}$ denotes how much planned capacity is still available (as a fraction of the planned demand).
\begin{equation}\label{eq:resy}
    \res{\Yreq} \triangleq 
    \Bigl\{ \yreq{\sapp}{\snet} \rdreq* - 
    \smashoperator{\sum_{\req \in \rreq \cap \planreqs \cap \treqs}}
    \xreq{\sapp}{\snet} \dreq*
    \Bigr\}_{\substack{\rreq\in\rreqs,\\ \sapp \in \rgreq, \mathrlap{\snet\in\gnet}}}
\end{equation}
The residual capacity and residual plan are dynamically updated after allocating a request in Function~\textproc{Allocate} (\cref{lin:twentyeight,lin:thirty}). 
We use \eref{eq:feasiS} and \eref{eq:feasiY} to check if a possible allocation $\Xreq[\req]$ is feasible w.r.t. the substrate and plan residuals.
$\res{\gnet} \ge 0$ if all its elements are non negative; similarly for $\res{\Yreq} \ge 0$.
\begin{align}\label{eq:feasiS}
    \Xreq[\req] &\le \res{\gnet} \text{ if } \res[t, \Xreq \cup \set{\Xreq[\req]}]{\gnet} \ge 0
\\\label{eq:feasiY}
    \Xreq[\req] &\le \res{\Yreq} \text{ if } \res[t, \Xreq \cup \set{\Xreq[\req]}]{\Yreq} \ge 0
\end{align}
\eref{eq:feasiS} means that it is feasible, under the capacity constraints of $\gnet$ to add $\Xreq[\req]$ to $\Xreq$.
Similarly, \eref{eq:feasiY} means that $\Xreq[\req]$ can be (unsplittably) allocated from $\Yreq$, considering the demand of all other processed allocations that followed the plan $\Yreq$. 

Upon the arrival of each request $\req$ (\cref{lin:newreq}), \OLIVE first examines planned candidate embeddings (Function~\textproc{PlanEmbed}). If the residual plan indicates that $\req$ can be allocated according to $\Yreq$ then $\req$ is marked as \textit{planned=True} (\cref{lin:plannedok}). The plan is already optimized for the most cost-efficient embedding of the aggregate demand. Thus, any embedding that follows the plan is good and there is no need to optimize further for cost. If, however, $\req$ cannot be allocated according to $\Yreq$, then $\req$ is marked as \textit{planned=False} and an embedding must be found ad-hoc using the \textit{non-planned mechanisms} to compensate for the deviation from the plan. 

Non-planned allocations ``borrow'' unused capacity from $\Yreq$, thus subsequently arriving requests might not be able to follow their planned allocation. To overcome this, when \OLIVE detects that there is not enough residual substrate capacity to allow a planned embedding (\cref{lin:premept}) it \textit{preempts} requests to free the ``borrowed'' resources (Function~\textproc{Preempt}). The preempted requests incur the rejection cost. 

\OLIVE applies two non-planned embedding mechanisms. First, it attempts to allocate any fraction of $\req$ according to $\Yreq$ (\cref{lin:thirtyeight}), subject to feasibility constraints w.r.t. the residual capacity. The resulting allocation $\Xreq$ overflows the plan $\Yreq$, ``borrowing'' unused capacity. While this allocation may not be fully optimized, it is preferred over not using the plan at all, as every fractional allocation that follows the plan already considers placement preferences and expected overall demand. If $\Xreq$ is not feasible, another candidate embedding must be found (\cref{lin:nox}). In this case, \OLIVE uses a \textit{Greedy} approach (Function~\textproc{GreedyEmbed}), selecting a feasible embedding for $\req$ with the lowest resource cost. 

\textproc{GreedyEmbed} finds a candidate embedding for $\req$ by solving an \ofvne problem, setting $\res{\gnet}$ as the substrate capacities. Since finding an exact solution might not scale well, \textproc{GreedyEmbed} restricts the embedding, such that all virtual nodes are collocated on the same substrate node. With this heuristic restriction, a shortest-path calculation is used to efficiently find the least-cost candidate embedding~\cite{Rexford2008,DeepVine}. This approach may also improve solution quality by reducing link congestion~\cite{rubio2024novel}.

\section{Evaluation}\label{sec:eval}


\newcommand{\figrejall}
{\begin{figure*}
\captionsetup[subfloat]{justification=centering,format=hang,captionskip=1pt}%
    \centering%
    \includegraphics[scale=0.4,trim=2.1cm 2.1cm 7.4cm 2.3cm,clip]{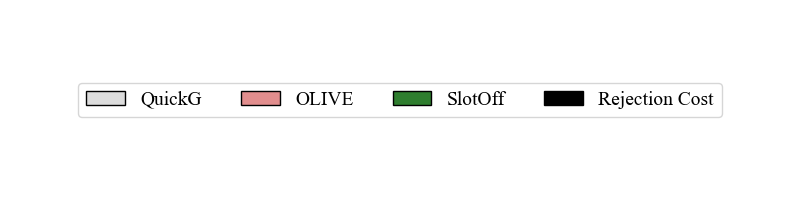}
    \\[-1.1\baselineskip]
    \hfill\centering
    \subfloat[Iris]{{\includegraphics[width=0.24\linewidth,trim=0cm 0 0cm 0cm,clip]{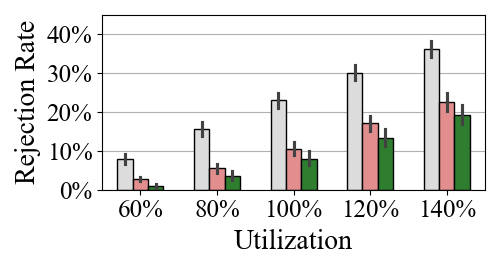}}\label{fig:iris_rej}}
    \hfill
    \subfloat[Citta Studi]{\includegraphics[width=0.24\linewidth,trim=0cm 0 0cm 0cm,clip]{fig/Rejection_Rate_Iris_full_2_accum.png}}
    \hfill
    \subfloat[5GEN]{\includegraphics[width=0.24\linewidth,trim=0cm 0 0cm 0cm,clip]{fig/Rejection_Rate_Iris_full_2_accum.png}}
    \hfill
    \subfloat[100N150E]{\includegraphics[width=0.24\linewidth,trim=0cm 0 0cm 0cm,clip]{fig/Rejection_Rate_Iris_full_2_accum.png}}
    \vspace{-1mm}
    \caption{Rejection rate}
    \vspace{-2mm}
    \label{fig:all_rej}
\end{figure*}}

\newcommand{\figcostall}
{\begin{figure*}
    \captionsetup[subfloat]{justification=centering,format=hang,captionskip=1pt}%
    \centering%
    \includegraphics[scale=0.4,trim=2.1cm 2.1cm 2.1cm 2.3cm,clip]{fig/legend_cost.png}
    \\[-1.1\baselineskip]
    \hfill\centering
    \subfloat[Iris]{{\includegraphics[width=0.24\linewidth,trim=0cm 0 0cm 0cm,clip]{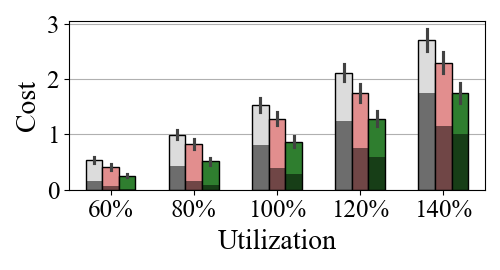}}\label{fig:iris_cost}}
    \hfill
    \subfloat[Citta Studi]{\includegraphics[width=0.24\linewidth,trim=0cm 0 0cm 0cm,clip]{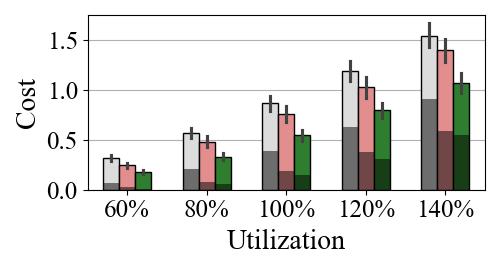}}
    \hfill
    \subfloat[5GEN]{\includegraphics[width=0.24\linewidth,trim=0cm 0 0cm 0cm,clip]{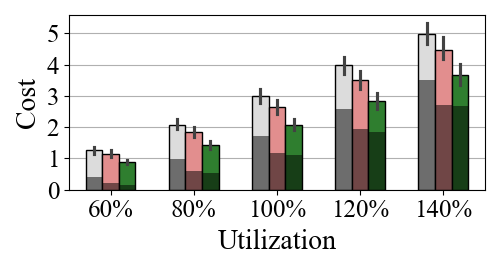}}
    \hfill
    \subfloat[100N150E]{\includegraphics[width=0.24\linewidth,trim=0cm 0 0cm 0cm,clip]{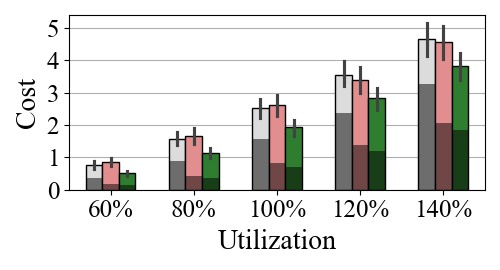}}
    \vspace{-3mm}
    \caption{Cost}
    \vspace{-4mm}
    \label{fig:all_cost}
\end{figure*}}

\newcommand{\figtopoall}
{\begin{figure*}
    \captionsetup[subfloat]{justification=centering,format=hang,captionskip=1pt}%
    \centering
    \hspace{.03\linewidth}
    \subfloat[Iris]{\includegraphics[width=0.18\linewidth,trim=0cm 1cm 0cm 1cm,clip]{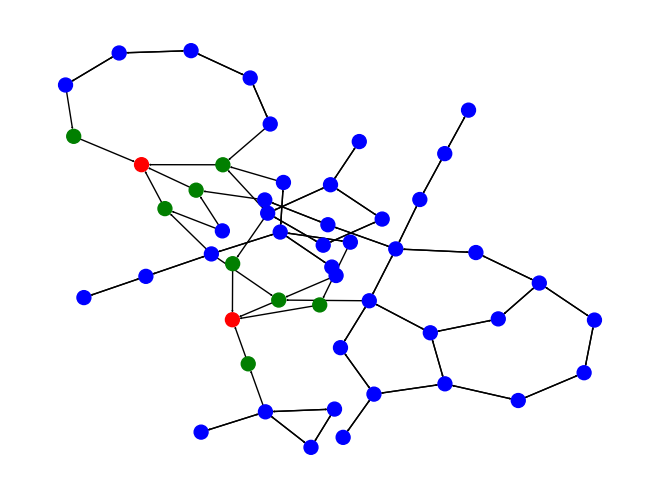}}
    \hfill
    \subfloat[Citta Studi]{\includegraphics[width=0.18\linewidth,trim=0cm 1cm 0cm 1cm,clip]{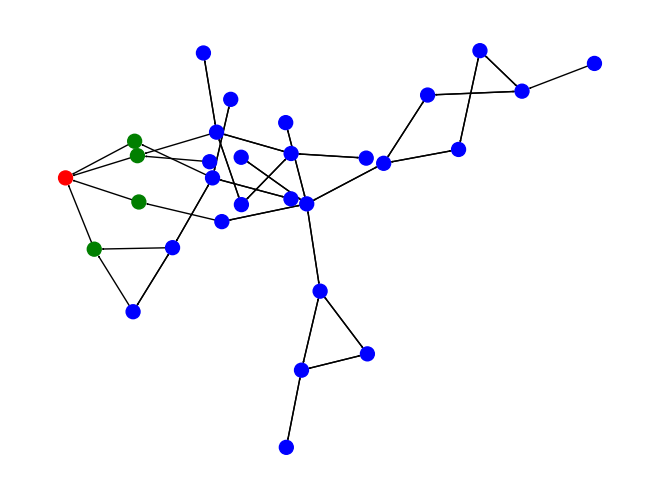}}
    \hfill
    \subfloat[5GEN]{\includegraphics[width=0.18\linewidth,trim=0cm 1cm 0cm 1cm,clip]{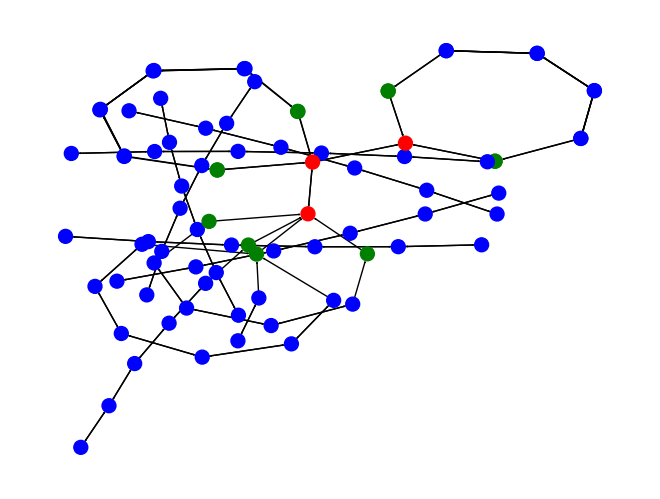}}
    \hfill
    \subfloat[100N150E]{\includegraphics[width=0.18\linewidth,trim=0cm 1cm 0cm 1cm,clip]{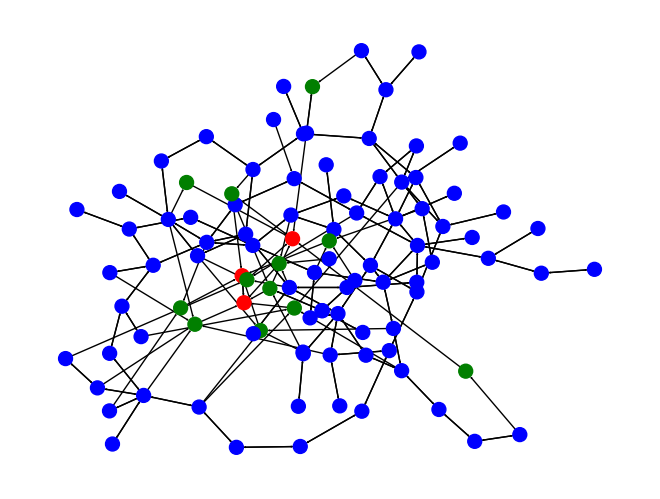}}
    \hspace{.02\linewidth}
    \caption{Physical Topologies. Edge, transport, and core \dcs are represented by blue, green, and red, respectively.}
    \label{fig:all_topo}
    \vspace{-2mm}
\end{figure*}}

\newcommand{\figdemandbyt}
{\begin{figure}
    \centering
        \includegraphics[scale=0.35]{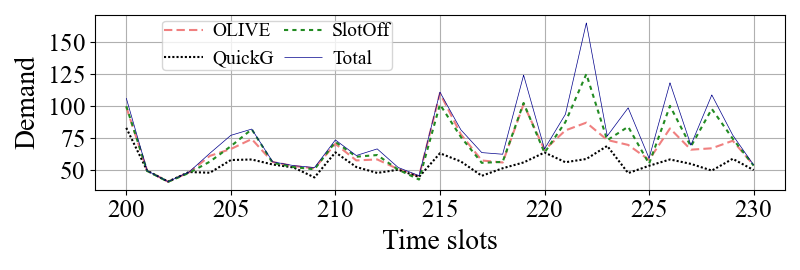}
    \caption{Iris, 140\% util.: allocated demand per time slot; slots: 200-250}
    \label{fig:demand_by_t}
\end{figure}}

\newcommand{\difftraindemand}
{\begin{figure}
    \begin{minipage}[b]{0.4\linewidth}
        \raggedright
    \begin{minipage}[b]{0.41\linewidth}
        \centering
    \includegraphics[scale=0.35]{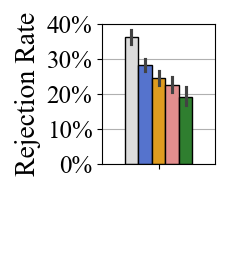}
    \end{minipage}\noindent
        \begin{minipage}[b]{0.12\linewidth}
        \centering
        \includegraphics[scale=0.35,trim=7.6cm 0cm 0cm 0cm,clip]{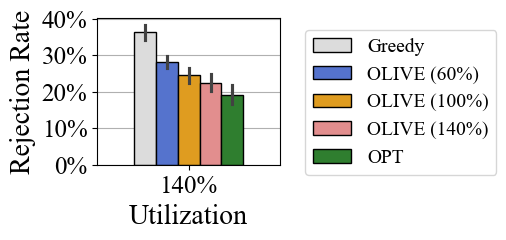}
    \end{minipage}
    
    \caption{\OLIVE plan utilization of 60\%, 100\%, and 140\%, test with utilization 140\%, Iris.}
    \label{fig:iris_train_dem_140}
    \end{minipage}
\end{figure}}

\newcommand{\figeightnine}
{\begin{figure*}\centering
    \begin{minipage}[b]{0.39\linewidth}
        \centering
        \includegraphics*[trim=0.35cm 0.2cm 0.37cm 0cm, scale=0.35]{fig/Allocated_Arrived_Demand_Iris_demand_plot_14_2_line_plot.png}
    \caption{Zoom in on time slots 200-230.}
    \label{fig:demand_by_t}
    \end{minipage}%
\hfill\begin{minipage}[b]{0.24\linewidth}
        \centering
        \includegraphics*[trim=0.35cm 0.2cm 0.37cm 0cm, scale=0.35]{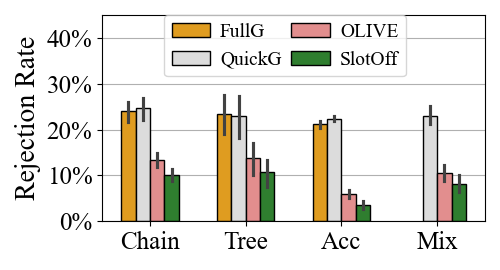}
        \caption{Iris: Rejection rate by\\application type.}
        \label{fig:apps}
    \end{minipage}%
\hfill\begin{minipage}[b]{0.15\linewidth}
        \centering
        \begin{tabular}[t]{@{}c@{}c@{}}
        \includegraphics*[trim=0.35cm 0.2cm 0.35cm 0cm, scale=0.35]{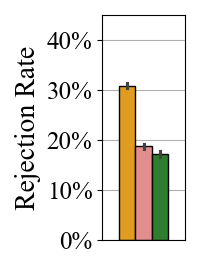} &
        \includegraphics*[trim=8.6cm 0cm 8.5cm 1cm,scale=0.35]{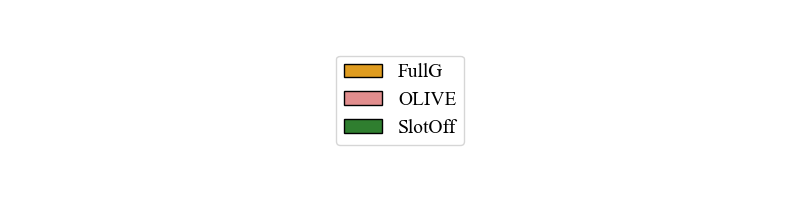}
        \end{tabular}
        \caption{Iris: GPU.}
        \label{fig:gpu}
    \end{minipage}%
\hfill\begin{minipage}[b]{0.14\linewidth}
        \centering
        \includegraphics*[trim=0.35cm 0.2cm 0.37cm 0cm, scale=0.35]{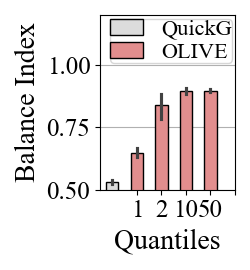}
        \caption{Balance\\index by \layertxts.}
        \label{fig:layers}
    \end{minipage}%
\end{figure*}}

\newcommand{\figshiftreal}
{\begin{figure*}
    \begin{minipage}[t]{.2\linewidth}\vspace{9pt}
        \begin{tabular}{@{}c@{}c@{}}
            \includegraphics[scale=0.35,trim=0.2cm 0cm 0cm 0cm,clip]{fig/Rejection_Rate_Iris_diff_train_test_dem_0_accum.png} & 
            \raisebox{0.68cm}{\includegraphics[scale=0.35,trim=7.7cm 0cm 7.5cm 0cm,clip]{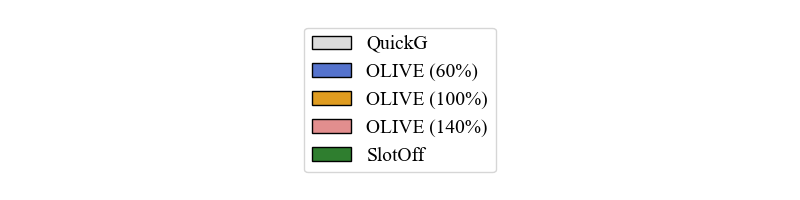}} \\
        \end{tabular}
        \vspace{2pt}
        \caption{Effect of deviation from plan.}
        \label{fig:iris_train_dem_140}
    \end{minipage}%
    \hfill
    \begin{minipage}[t]{0.77\linewidth}\vspace{0pt}
        \centering%
        \includegraphics*[trim=0cm 3cm 0cm 3cm,scale=0.33]{fig/legend_cost.png}\\
        \begin{minipage}[b]{0.5\linewidth}\vspace{0pt}%
            \centering%
            \subfloat[Rejection rate]{%
                \includegraphics[scale=0.35,trim=0 0 0cm 0,clip]{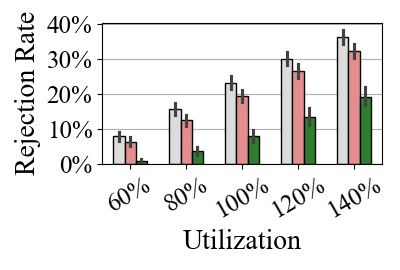}%
                \label{fig:iris_shif_rej}%
            }%
            \subfloat[Cost]{%
                \includegraphics[scale=0.35,trim=1cm 0 0cm 0,clip]{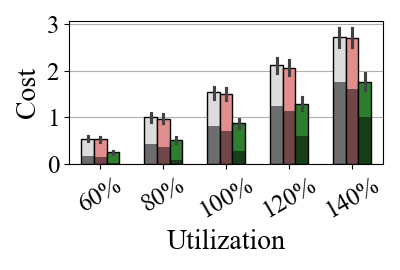}%
                \label{fig:iris_shif_cost}
            }%
            \caption{Shifted plan requests in Iris.}
            \label{fig:iris_shifted_plan}%
        \end{minipage}
        \begin{minipage}[b]{0.5\linewidth}\vspace{0pt}%
            \centering%
            \subfloat[Rejection rate]{%
                \includegraphics[scale=0.35,trim=0 0 0cm 0,clip]{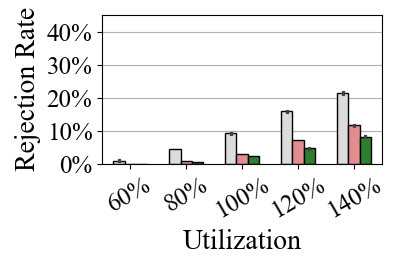}%
                \label{fig:iris_real_rej}%
            }%
            \subfloat[Cost]{%
                \includegraphics[scale=0.35,trim=1cm 0 0cm 0,clip]{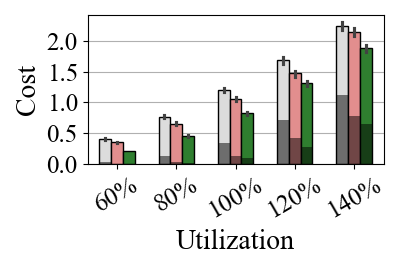}%
                \label{fig:iris_real_cost}%
            }%
            \caption{Real demand in Iris.}
            \label{fig:iris_real}
        \end{minipage}
    \end{minipage}
    \vspace{-3mm}
\end{figure*}}

\newcommand{\runtime}
{\begin{figure*}[t!]
\captionsetup[subfloat]{justification=centering,format=hang,captionskip=1pt}%
\centering
    \includegraphics*[trim=0cm 2.1cm 0cm 2.1cm,scale=0.36]{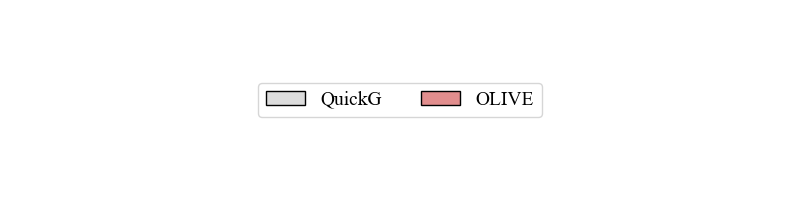}
    \\[-.7\baselineskip]
    \centering
    \subfloat[Iris, different arrival rate]{{\includegraphics[scale=0.35,trim=0cm 0 0cm 0.2cm,clip]{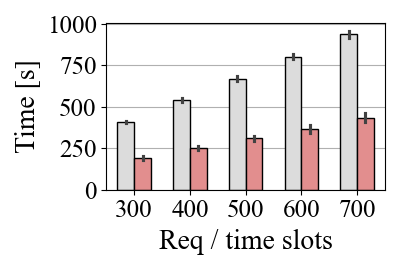}}\label{fig:runtime_requests}}
    \hfill
    \subfloat[Iris]{\includegraphics[scale=0.35,trim=0cm 0 0cm 0.2cm,clip]{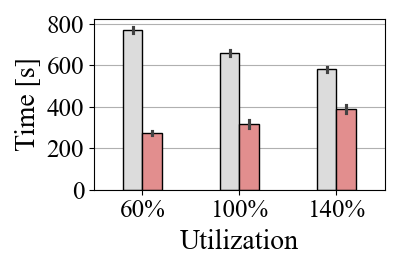}}
    \hfill
    \subfloat[Citta Studi]{\includegraphics[scale=0.35,trim=0cm 0 0cm 0.2cm,clip]{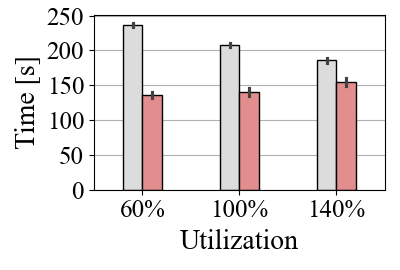}}
    \hfill
    \subfloat[5GEN]{\includegraphics[scale=0.35,trim=0cm 0 0cm 0.2cm,clip]{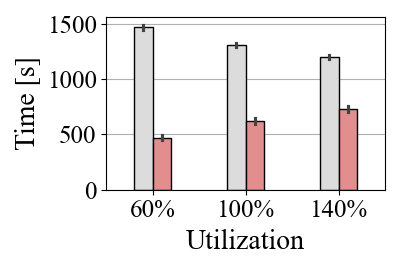}}
        \hfill
    \subfloat[100N150E]{\includegraphics[scale=0.35,trim=0cm 0 0cm 0.2cm,clip]{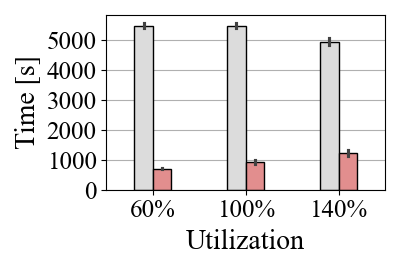}}
    \caption{Runtime scalability for request arrival rate and topology size}
    \label{fig:runtime}
    \vspace{-3mm}
\end{figure*}}


\ignore{Because of the solutions and approaches proliferation reported in \cref{sec:related}, it is infeasible to compare \OLIVE directly with all of them \OK{perhaps rephrase}. Instead, we compare our solution to the theoretical optimum and greedy strategy that jointly optimizes mapping of virtual nodes and links. 
The greedy approach, common in many heuristic solutions, does not involve any planning.}


In our experimental evaluation, we explore the benefit attained by combining globally optimized offline planning with dynamic mechanisms of capacity re-balancing when using the plan at run time. In addition, we evaluate the sensitivity of \OLIVE to the network topology, the distribution of load on the substrate, virtual application topologies, and differences between the expected and encountered demands.

\subsection{Experimental Setup}\label{sec:eval-setup}

\T{Algorithms} It is infeasible to compare \OLIVE to the myriad of existing solutions (see \cref{sec:related}). The challenges include non-repeatability due to the lack of open-source implementations and/or datasets, as well as insufficient details on algorithm fine-tuning in existing solutions.
To that end, we consider three algorithms that represent the state-of-the-art, are well-documented, and are straightforward to reproduce. 
\Greedy runs \OLIVE with an empty plan, resorting to greedily allocating each request, applying the heuristic approach of 
\textproc{GreedyEmbed} (\cref{sec:olive}). 
\GreedyILP is similar to \Greedy, but omits the collocation restriction of VNFs of the same request. It solves a much harder \ofvne instance for each request, utilizing an ILP solver to find an exact solution. 
\GreedyILP is the best possible greedy algorithm, but it does not scale well. It is not intended to be used in practice and is evaluated only as a reference point. 
\ignore{The first one is  \Greedy, the greedy approach described in \cref{sec:solution}. \GreedyILP is a greedy algorithm that solves \pvne exactly for each individual request arriving online. It consumes a lot of resources and its running time is long. \GreedyILP is not intended to be used in practice. We refer to it because this is the best possible greedy algorithm. }\ignore{It represents the best online policy which considers each embedding request independently. }
\ignore{
\OPT serves as a lower bound on the optimal allocation for each time slot. It is obtained by solving the aggregated \DB{we do not need to say "aggregated" here} \ofvne for all active \DB{no need to say "active"} requests within each time slot and then heuristically dividing the aggregated demand among individual requests by serving requests in a first-come-first-serve approach.\DB{I think this means that LowB is \textbf{not} the fractional optimum, because we round it. The only thing that is important here is that we use actual demand that happens in this time slot (i.e., we have the full knowledge of the slot rather than discovering requests one by one). We need not refer \ofvne, which is now not defined explicitly. However, we can still say offline \vne. BTW, why don't we simply say here that we solve it using~\cite{pranos} for a single slot if we do not stop at LP but also round? If agreed, I'd like to rephrase it and also references to it in the earlier parts of the paper.}\DL{I don't think the definition of \OPT is clear from the text, let alone its optimality as a lower bound. I like David's idea of phrasing this as utilizing \textsc{Pranos}. \\
Intuitively, I think of \OPT as breaking each request into 1-slot intervals, allowing reallocation of a request in each time slot. We then solve \ofvne for each time slot. The main reason why this is not a real lower-bound is that we do not reconsider already rejected requests in subsequent time slots. \\
\textbf{Suggested rephrasing:}\\
}
}
\OPT sequentially computes an allocation for each time slot $t$, by solving a separate \ofvne instance comprising of the active requests, $\treqs$. Note that ongoing active requests may have a completely different allocation for each time slot.\footnote{Giving \OPT an inherent advantage when compared to \OLIVE.} However, rejected requests are not reconsidered in the subsequent time slots. 
Our \OPT uses \textsc{Pranos}~\cite{pranos} for solving \ofvne per time slot, as it produces near-optimal solutions and is highly scalable.
\ignore{
While any offline \vne algorithm can be used for solving \ofvne per time slot, they either do not scale well or are susceptible to local minima or overly sensitive to data or model parameters, as discussed in~\ref{sec:intro}. To that end, in \OPT, we selected \textsc{Pranos}~\cite{pranos} as a building block that represents SoTA for \ofvne and produces near-optimal solutions for the large scale \ofvne problems. }
\ignore{
\OK{not sure the reader will understand why this is near optimal. Clearly state (remind) PRANOs is state-of-the-art near optimal. ``allowing reallocation of requests'' -- should we mention OLIVE is not allowed to do it?} \OK{Do we remove the next sentence?}
If a request is rejected in any of the time slots during which it is active, it is rejected for all time slots. \OPT is considered a lower bound because each virtual node of a request is allowed to be embedded onto different physical nodes across subsequent time slots.\DB{We need to decide whether \OPT is fractional as stated here or not as stated in the previous paragraph :)}\DB{If we are Ok with the rephrasing suggested in previous comment, let's remove most of this description.} Ideally, \pvne, with all individual requests as input, would be used to derive the true optimal solution; however, solving such a large MILP within a reasonable timeframe is impractical.\DB{Once we decide whether we use LP or rounded LP, we need to revisit this last sentence.}
\OK{I don't think this sentence is correct since \pvne is now offline.}
}

\T{Physical topologies} We used topologies from four sources: (1) \textbf{Iris}, a realistic topology from the Internet Topology Zoo~\cite{zoo}, a collection of network topologies commonly used for evaluation~\cite{tzoo1,tzoo2}, (2) \textbf{Citta Studi}, a realistic mobile edge network topology~\cite{pranos,citta1}, (3) \textbf{5GEN}, a topology representing 5G network deployment in Madrid, Spain, created by the 5GEN tool~\cite{5gen}, and (4) \textbf{100N150E}, a large connected  Erd\H{o}s-R\'enyi random graph~\cite{pranos,citta1}. The topologies are depicted in \cref{fig:all_topo}.

In alignment with the mobile access network architecture, we consider three tiers of nodes: \textit{edge}, \textit{transport}, and \textit{core}~\cite{pranos,5gpp}. For each pair of successive tiers (e.g., edge and transport), we set a ratio of 3 between their link capacities and \dc capacities, as in~\cite{chiang2020virtual}. 
Datacenter and link costs were assigned similarly to~\cite{pranos}, with \dc costs uniformly distributed between 50\% and 150\% of the mean \dc cost in each tier. We use generic \textit{capacity units} (\textsc{CU}) for node and link capacities, similar to the definition in~\cite{pranos}.
\Cref{tab:topologies} summarizes  the physical topologies we use.

\T{Virtual network} 
We examine four application topologies: a \textit{chain}, a \textit{tree} with two branches, an \textit{accelerator}, and a \textit{GPU}. The accelerator application is a chain with a single \textit{accelerator function}, which reduces the size of the consequent virtual link by 70\%, similar to the application in~\cite{forti2022probabilistic}. The GPU application is a chain with a randomly selected \textit{GPU VNF} that must be placed on dedicated \textit{GPU \dcs}. These \dcs, which do not allow placement of non GPU VNFs, are designated by setting the coefficients $\dnet{\sapp}{\snet}$ (see \cref{sec:preliminaries}).
We use four application instances drawn with equal probabilities: two chain applications, a tree application, and an accelerator application. 
The number of VNFs per topology is distributed uniformly between three to five instances. We also tested virtual networks of different sizes and structures and obtained similar results, omitted due to space considerations. Virtual node and link sizes follow a normal distribution with a mean of 50 and a standard deviation of 30. 

\T{Traces} We use two traces of 6000 time slots. The first 5400 slots are used to form $\histreqs$, and the remaining 600 slots are used for the online phase. 
In both traces, request demands follow a normal distribution, $\mathcal{N}(10,4)$ with requests exclusively originating from the edge \dcs, as expected in a realistic scenario. Request duration followed an exponential distribution with a mean of 10 time slots. \Cref{tab:settings} summarizes the parameters of our experimental settings.

The first trace is generated using a Markov-modulated Poisson process (MMPP)~\cite{miao2021performance}. MMPP consists of high ($\lambda_h$) and low ($\lambda_l$) arrival-rate states with a Markov transition process between them and mean $\lambda$ set to 10 per node in our experiments. For example, in the 100N150E topology with 100 nodes, this results in an average of $1000$ requests per time slot, with momentary demand bursts. The use of MMPP effectively captures the bursty nature of realistic request arrivals~\cite{miao2021performance, muscariello2005markov}. 
The second trace is derived from the “Equinix-NewYork” network monitor in 2019 CAIDA Internet traces~\cite{CAIDA_trace}. Similarly to~\cite{CAIDA}, we aggregate requests from the same IP sources and randomly assigned the grouped requests to the \dcs. \ignore{We multiply the timestamp of each request by 100 to extend the traces to 6000 time slots. }Adapting Internet traces to an edge setup is necessary due to the absence of realistic mobile access network workloads~\cite{kolosov2020benchmarking}.

\T{Methodology} In all experiments, we execute \OLIVE, \Greedy, \GreedyILP, and \OPT 30 times on the four topologies. 
In each execution, we draw an  application set from the distribution given in \cref{tab:settings}. We present the averages and confidence intervals for the rejection rate and cost. We say that the \textit{edge utilization} is 100\% when the mean size of all active requests equals the total capacity of all edge \dcs in a specific topology. We examined utilization values between 60\% and 140\% by changing the mean request demand size within the range of 6--14. We display the results for requests that started between time slots 100 and 500.

\T{Execution environment} 
Our algorithms were implemented in Python 3.10, using CPLEX 22.1.1~\cite{CPLEX} to solve the LP of \OPT and \OLIVE planning and the ILP of \GreedyILP. 
Experiments were executed on an Ubuntu 18.04 server with an Intel Xeon Gold 6458Q CPU (3.1 GHz) and 500GB of RAM.

\begin{table}
\centering
\footnotesize
\caption{Details of the Topologies}
\begin{tabularx}{.9\linewidth}{@{}>{\hsize=.5\hsize\linewidth=\hsize}Xcc>{\hsize=1.5\hsize\linewidth=\hsize}X@{}}
\toprule
\textbf{Topology} & \textbf{Nodes} & \textbf{Links} & \textbf{Description} \\ \midrule
Iris & 50 & 64 & Topology Zoo \cite{zoo} \\ 
Citta Studi & 30 & 35 & Edge network topology \cite{citta1,pranos} \\ 
5GEN & 78 & 100 & Realistic 5G topology \cite{5gen,pranos} \\ 
100N150E & 100 & 150 & Random graph \cite{citta1,pranos} \\ 
\bottomrule
\end{tabularx}

\vspace{0.5cm} 
\begin{tabularx}{.9\linewidth}{@{}Xccc@{}}
\toprule
\textbf{Parameter} & \textbf{Edge} & \textbf{Transport} & \textbf{Core} \\ \midrule
Node Cap [\textsc{CU}] & 200K & 600K & 1.8M \\ 
Mean Node Cost (per \textsc{CU}) & 50 & 10 & 1 \\ 
Link Cap [\textsc{CU}] & 100K & 300K & 900K \\ 
Link Cost (per \textsc{CU}) & 1 & 1 & 1 \\ 
\bottomrule
\end{tabularx}
\label{tab:topologies}
\vspace{-3mm}
\end{table}

\begin{table}
\centering
\footnotesize
\caption{Experimental Settings}
\begin{tabularx}{.9\linewidth}{@{}XX@{}}
\toprule
\textbf{Parameter} & \textbf{Value} \\ \midrule 
Node popularity & Zipf ($\alpha=1$)  \\ 
Plan period [time slots] & 5400  \\ 
Test period [time slots] & 600  \\ 
Request Size & $\mathcal{N}(10, 4)$\\ 
Request Duration & Exponential, mean: 10\\ 
Requests per node ($\lambda$) & 10 per time slot\\ 
Applications & 2 chain, 1 tree, 1 accelerator\\ 
VNFs & $\mathcal{U}(3, 5)$\\ 
Application function size & $\mathcal{N}(50, 900)$\\ 
Application link size & $\mathcal{N}(50, 900)$\\ 
\bottomrule
\end{tabularx}
\label{tab:settings}
\end{table}

\subsection{Results}

\figtopoall%
\figrejall%
\figcostall%


\T{Request rejection rate} \Cref{fig:all_rej} shows the rejection rate for each topology under utilization values between 60\% and 140\%. 
As expected, the rejection rate increases with the utilization for all algorithms and tested topologies, as higher utilization levels leave fewer optimization opportunities. 
However, \OLIVE rejection rates were significantly lower than those of \Greedy and were very close to those of \OPT with a maximum gap of 4\%.

\cref{fig:demand_by_t} zooms into the demand (scaled down by 100 for convenience of presentation) allocated by the different algorithms (compared to the total requested demand) in Iris at 140\% utilization during time slots 200--230.  
\Greedy fails to allocate a large portion of the demand even during mild bursts. In contrast, \OLIVE allocates almost as much demand as \OPT during the mild bursts. At medium bursts, the non-planned mechanisms of \OLIVE allow it to compensate for deviations from the plan. Only at very high or prolonged bursts, \OLIVE momentarily differs from \OPT by a factor of 2, but still outperforms \Greedy by a factor of 2.

\T{Request embedding cost} \cref{fig:all_cost} shows the total cost of embeddings of the same experiments. 
We set a very conservative rejection penalty factor, $\psi(\req)$, that equals the cost of allocating elements $\sapp$ of $\areq$ on the most expensive elements $\snet$.

\OLIVE outperforms \Greedy for all utilization levels and in all topologies. \cref{fig:all_cost} is very compelling as it shows that \OLIVE outperforms \Greedy in terms of rejection rate by a factor of $2$ while incurring almost identical rejection cost as \OPT.

\ignore{
Recall that \OLIVE does not minimize its embedding cost as a primary goal. Rather, it uses the min-cost plan to minimize its rejection cost in the online embedding. Nevertheless, its cost was similar to that of \OPT in many scenarios. The difference between the cost of \OLIVE and that of \OPT increased with the utilization, due to two reasons: the increase in the rejection rate of \OLIVE (with respect to that of \OPT), and the increasing deviation from the plan. \OLIVE outperformed \Greedy in most scenarios. }

\T{Effect of individual applications} \ignore{Next, we examine the sensitivity of the rejection rate to application types.} 
\cref{fig:apps} shows the sensitivity of the rejection rate to application types in Iris at 100\% utilization. In each experiment, we used four applications of the same type with parameters distributed as specified in \cref{tab:settings}. The results of the application mix from \cref{fig:iris_rej} are included for reference. 

\Greedy is not sensitive to the application type and achieves similar rejection rates in all cases.
Also, as expected at this load level, \GreedyILP and \Greedy achieve similar results statistically (as evidenced by the confidence intervals shown in the graph).
\ignore{With the tree applications, the rejection rate of \GreedyILP is higher than that of \Greedy due to unnecessary fragmented allocations, which were shown to increase rejection rates~\cite{rubio2024novel}.} However, the runtime of \GreedyILP was $130\times$ longer than that of \Greedy. Thus, except in \cref{fig:gpu}, we use only \Greedy for the rest of the evaluation.
In contrast to \Greedy, the rejection rate of \OLIVE was significantly lower and much closer to that of \OPT. As expected, it decreases with the introduction of the accelerator function (`Acc’ and `Mix’). 

\figeightnine

\begin{figure}
    \centering
    \includegraphics[width=1\linewidth]{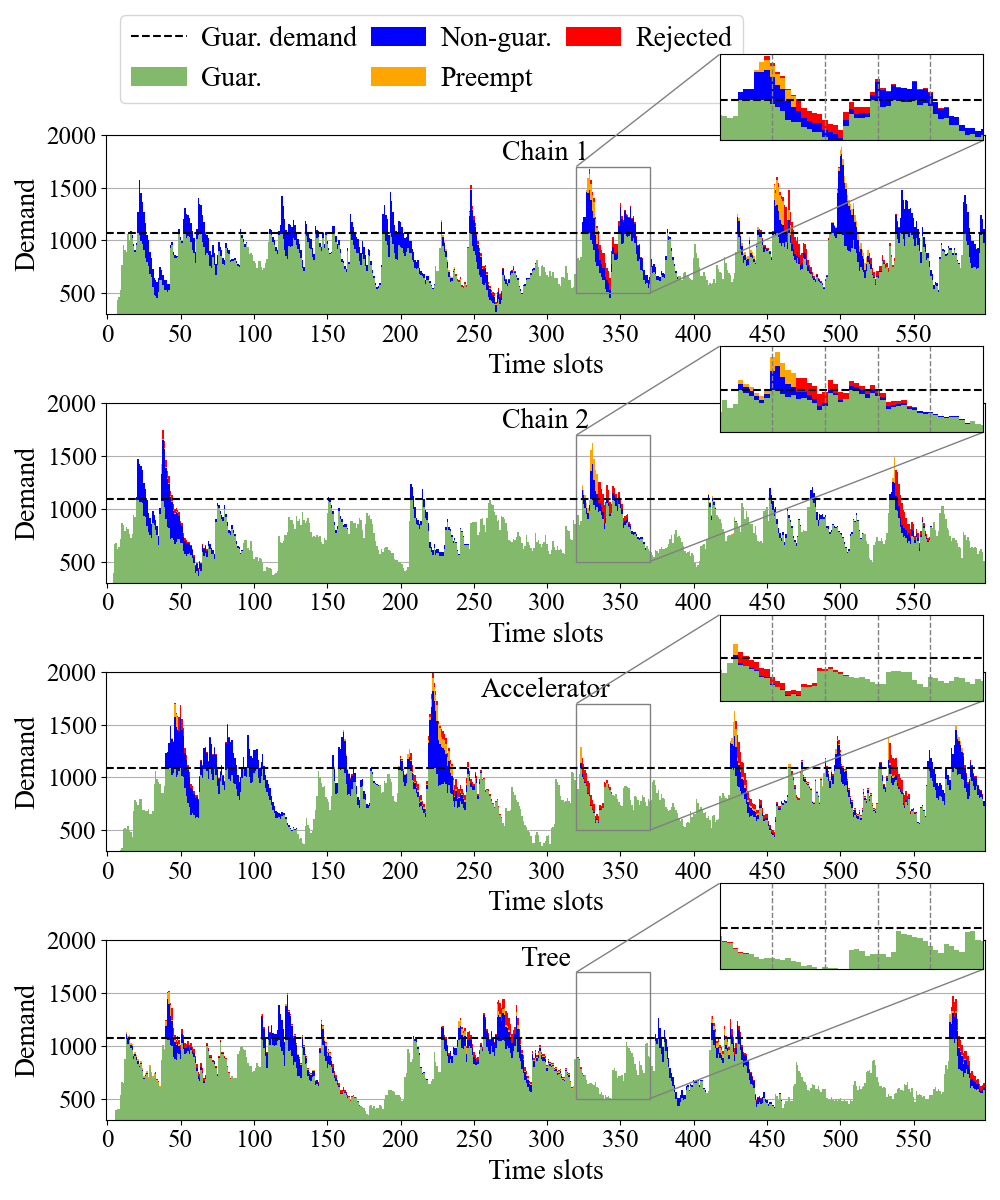}
    \vspace{-6mm}\caption{Franklin node (Iris, MMPP), \OLIVE guaranteed demand (horizontal line) compared to actual allocation.}
    \label{fig:franklin}
    \vspace{-4mm}
\end{figure}

\figshiftreal
\runtime

\Cref{fig:gpu} shows the rejection rates of \GreedyILP, \OLIVE, and \OPT, for a GPU scenario. We repeat the experiment in \cref{fig:iris_rej} for 100\% utilization with four chain applications, each containing one \textit{GPU function}. 
We modify Iris to support this scenario by splitting the core nodes and four random edge nodes into GPU and non-GPU ones. Non-GPU \dcs were assigned capacity smaller by 25\%. The rejection rate of \OLIVE was only 2\% higher than that of \OPT and 12\% lower than that of \GreedyILP. Though the GPU allocation constraint limits the solution space, \OLIVE still significantly outperformed \GreedyILP, demonstrating its flexibility in complex allocation tasks. \Greedy was not evaluated in this experiment due to its co-location requirement, i.e., that all virtual nodes are collocated on the same substrate node.

\T{Balanced allocation}
We define a \textit{rejection balance index} to measure how rejections are distributed between different request classes. Similar to Jain's index~\cite{jain}, we define it as: 
\begin{equation}\label{eq:rr}
\frac{1}{\sum_{v \in \gnet}n(v)}\sum_{v \in \gnet} \frac{n(v)(\sum_{\app \in \allapps}x_{va})^2}{\abs{\allapps}\cdot\sum_{\app \in \allapps}x_{va}^2}
\end{equation}
\ignore{where $n$ is the number of \dcs, $d_i$ is the number of requests from node $i$, $x_j$ is the number of rejected requests of application $j$, and $a$ is the number of applications (4 in our experiment). To account for all \dcs, we calculated the weighted average using the total number of requests in each \dc as weights. Since applications have equal probabilities, we expect an equal number of rejected requests for each application in each \dc. A value of 1 indicates perfect balance. 
\DL{Maybe:\\} \OK{Yes, this version is correct}
\begin{equation}\label{eq:rr}
\frac{1}{\sum_{v \in \gnet}n(v)}\sum_{v \in \gnet} \frac{n(v)(\sum_{\app \in \allapps}x_{va})^2}{\abs{\allapps}\cdot\sum_{\app \in \allapps}x_{va}^2}
\end{equation}}
where $n(v)$ is the number of requests at \dc $v$ and $x_{va}$ is the number of rejected requests at $v$ of application $a$. We calculate a weighted average using $n(v)$ as weights. Since applications have equal probabilities, for each $v$, we expect $x_{va}$ to be similar for all applications. A value of 1 indicates a perfect balance.

We explore the effect of rejection \layertxts (\cref{sec:planvne}) on the rejection balance index using an extreme scenario. We used the same traces as \cref{fig:iris_rej} at 140\% utilization (which resulted in a large number of rejections). 
\cref{fig:layers} shows the rejection balance index for \Greedy (i.e., with no \layertxts) and for \OLIVE with 1, 2, 10, and 50 \layertxts. In \OLIVE the index was 0.65 for one \layertxt and 0.84 for two \layertxts. For 10 and 50 \layertxts the index was 0.89. \Greedy cannot actively balance between rejected applications. Consequently, its index was 0.53.  The result shows that planning with \layertxts improves the balance of rejection rate across different request classes. \Cref{fig:layers} also shows that increasing \layertxts beyond 10 did not yield further improvement. Therefore, we used 10 \layertxts in all experiments.

To obtain a deeper insight, we zoom in on individual nodes during the experiments of \cref{fig:iris_rej}. For example, on the allocation of requests at the `Franklin' node in Iris during a single execution of the experiment in \cref{fig:iris_rej}  at 100\% utilization.  
\cref{fig:franklin} shows all active requests with each application at the `Franklin' node in Iris in a separate subfigure. The horizontal dashed line indicates the guaranteed (planned) demand, which is identical for each application. Requests that arrive when the demand is below this threshold are marked as `guaranteed' (green). Non-guaranteed requests (blue) are those that arrive when the demand is above the threshold. These are allocated by `borrowing' resources guaranteed to other applications. Non-guaranteed requests that did not complete their execution are marked in yellow, and their color changes to red upon preemption. The requests shown in red are rejected immediately upon arrival.

We take a closer look at time slots 320--370. Each 10 time slots are marked by a vertical line in the zoomed-in figure. In the first 10 time slots, the top three applications generated demand above the threshold, while the bottom application (tree) did not utilize all its guaranteed demand. This allowed the first two applications to `borrow' unused demand for non-guaranteed allocations. In the following 10 time slots, some of these allocations were preempted. In time slots 350--370, the top application (chain 1) experienced another burst. All its requests exceeding the threshold were successfully embedded as non-guaranteed allocations by `borrowing' unused resources guaranteed for other applications.
These results demonstrate the flexibility of our compensatory mechanisms which deal with  momentary deviations from the planned demand.


\T{Unexpected demand} To evaluate the effect of large deviations from the expected demand, we repeat the experiment in \cref{fig:iris_rej} at 140\% utilization, where the planning phase is based on expected utilization values of 60\% and 100\%. We denote these executions as \OLIVE (60\%) and \OLIVE (100\%), respectively. \cref{fig:iris_train_dem_140} shows the resulting rejection rates, with the results of \OPT, \Greedy, and \OLIVE from the corresponding execution added for reference. 

\OLIVE (60\%) and \OLIVE (100\%) achieved rejection rates higher by only 6\% and 3\% than \OLIVE (140\%), respectively. This shows that \OLIVE maintains its benefit from planning, even when it encounters demand which is significantly higher than expected. It also maintains its advantage over \Greedy: the rejection rates of \OLIVE (60\%) and \OLIVE (100\%) were 8\% and 4\% lower than those of \Greedy, respectively.

\T{Spatial distribution change} \cref{fig:iris_shifted_plan} shows the results of another variation of the experiment from \cref{fig:iris_rej}. Here, in the input for the plan, 
we replaced each request's \dc with a different random \dc. Nevertheless, the rejection rate of \OLIVE was still lower than that of \Greedy and both achieved similar costs. Together, the two experiments demonstrate that even in extreme scenarios, where the plan is created for the expected demand that is quite different from the observed one, the rejection rate of \OLIVE is never worse than that of \Greedy, and usually is significantly lower. 

\T{CAIDA Internet trace based demand} In \cref{fig:iris_real}, we repeat the experiment in \cref{fig:iris_rej} using the second trace described in~\ref{sec:eval-setup}. The average arrival rate is 495 requests per second. The request rejection rate of \OLIVE was similar to that of \OPT for utilization values between 60\% and 100\%. For higher values, the difference between these algorithms increased up to 4\%, similar to the behavior shown in \cref{fig:iris_rej}. The differences in cost (\cref{fig:iris_real_cost}) were smaller than in the synthetic workload (\cref{fig:iris_cost}), due to the different application set and trace characteristics. Nevertheless, the cost of \OLIVE was consistently lower than that of \Greedy for all utilization values.


\T{Runtime} We evaluated the simulation runtime of \OLIVE in two experiments, varying a different parameter in each one. In the first experiment, we varied the request arrival rate in Iris with 100\% utilization. We maintained the same utilization in all executions by scaling the mean request size. The results are presented in \cref{fig:runtime_requests}. As expected, the runtime of \OLIVE and \Greedy increased linearly with the increase in the arrival rate, as both process requests serially. 

In the second experiment, we fixed the request rate and varied the utilization of each topology between 60\% and 140\%. Figures \ref{fig:runtime}b-e show that the simulation runtime of \OLIVE was shorter by a factor of 1.7--7.8, 1.4--6.1, and 1.2--4.1, with utilization values of 60\%, 100\%, and 140\%, respectively. \OLIVE runtime increased with higher utilization because, at these levels, the residual plan is depleted more quickly, making it harder for the greedy search to find available paths.
In contrast, the runtime of \Greedy decreased as the utilization increased, due to its increased rejection rate. This is a side-effect of our implementation: \Greedy immediately rejects requests in a time slot if all \dcs are full. \OLIVE, on the other hand, first attempts to preempt requests that divert from the plan and to greedily allocate the request before it is rejected. Recall from \cref{fig:all_rej} that thanks to this approach \OLIVE achieves significantly lower rejection rates. Finally, we note \OLIVE's planning phase is also scalable, as \pvne is solved once and is independent of the number of requests~\cite{pranos}.



\section{Related Work}\label{sec:related}
\T{Exact solutions} Exact solutions to online \vne model the problem using an integer linear program or mixed-integer linear program~\cite{melo2013optimal,HeHZ-LCN2023,melo2015optimal}. Although they often achieve near-optimal solutions, they have limited scalability~\cite{ExactVNE-Survey-2016}.
PASE~\cite{PASE} presents an interesting approach in the context of user mobility across 5G/6G cells. The algorithm splits the problem into smaller sub-problems solved by ILP in parallel. \ignore{to online \vne in the context of user mobility is presented in PASE~\cite{PASE}. Multiple VN requests are speculatively embedded on a grid, based on an expected mobile user location. The model is solved under a configurable time limit and target optimality gap. To mitigate the scalability problem, they process requests in batches.}


\T{Heuristic solutions} Heuristic approaches address the scalability problems of exact solutions. The classic two-stage embedding heuristic ViNEYard~\cite{ChowdhuryRB-TON2012} first embeds virtual nodes by rounding a fractional LP solution, and then embeds virtual links on these nodes.  
In Cheng et al.~\cite{Cheng-VNE2011}, the substrate nodes are sorted by relative importance based on topological attributes. 
This approach was generalized by Cao et al.~\cite{CaoZY-JCN-2019}, where a single-stage heuristic simultaneously maps virtual nodes and links. 
Harutyunyan et al.~\cite{harutyunyan2021cost} consider re-optimization costs when a new request arrives, rather than considering residual capacity only. To the best of our knowledge, no heuristic-based approach solves the online \vne problem using an offline plan, as proposed in this work.  

\T{AI solutions} AI has gained popularity in addressing the online \vne problem~\cite{DVNE-DRL,DeepVine,wang2021proactive,gu2019deep,liu2023energy}. Lim et al.~\cite{LIM2023983} provide a systematic review of reinforcement learning and graph neural networks methods to solve online \vne.
Despite its potential, AI faces several key challenges. Among the more important ones is the model-to-reality gap: during the training phase, it is difficult to generate a representative training set that accurately reflects real VN embedding requests in real-time~\cite{wu2024ai}. 
As noted in~\cite{wu2024ai}, 
scalability remains a critical challenge, leading to sub-optimal performance and limited effectiveness in real-life online \vne scenarios.

\T{Metaheuristics-based solutions} Metaheuristics are extensively studied by the research community~\cite{Ruiz2018,Rubio-LoyolaACTMS,rubio2024novel}. The more difficult problem faced by these techniques is convergence time, which is sensitive to the initialization and configuration parameters of the guided search they employ.

\T{Relation to \ofvne} Our \onvne model shares several properties with the \ofvne model described in \textsc{Pranos}~\cite{pranos}. We adopt the same modeling for substrate, applications, requests, and costs. We also exploit the same similarity properties as part of request aggregation. In creating our offline plan, which requires solving an \ofvne problem, we incorporate some of the aggregation and relaxation techniques of \textsc{Pranos}. Like them, we were able to develop a highly-scalable algorithm that can process thousands and even millions of requests.

\T{Comparison on requests arrival rate} To the best of our knowledge, the highest request rate of $40$ embedding requests per second was reported in~\cite{DeepVine}. In contrast, we show scalability to $1000$ requests per second on large physical topologies, while producing near-optimal solutions. 
\section{Conclusions and Future Work}\label{sec:conclusions}


Online \vne is a central problem in edge network virtualization. Our novel approach addresses this challenge by combining the strengths of offline planning with dynamic online plan corrections. We evaluated \OLIVE through extensive simulations using a bursty synthetic trace and a CAIDA Internet-based trace, handling up to 1000 requests per time slot on realistic topologies from four different sources, with physical topologies of up to 100 nodes. 
This request arrival rate is two orders of magnitude higher than previous reports in the literature, while achieving near-optimal performance.


Thanks to their high expressiveness, the (in)efficiency coefficients in our formulation lend themselves to future extensions, such as succinct modeling of energy considerations. Our work presents new motivation for developing further specialized offline plans. These, for example, may account for \textit{time-dependent} expected demand. The modularity of our approach will allow it to use the planning mechanism best suited for each practical setting.
\ignore{Our results demonstrate that \OLIVE achieves near-optimal~\OK{revisit after we decide on \OPT} performance across all topologies, scaling efficiently with problem size and load while significantly outperforming a greedy strategies. \OK{Additional points: 1) Mention we intend to focus on latency and energy minimization in future work? (was raised in a TANTO review), 2) aggregation, bootstrap relaxation, rounding, are not new, but the novelty is in using them to solve a scalable \onvne}

\OK{Points to add: Planning is a known technique, but to the best of our knowledge, it was not applied to ON-VNE. Novelty is usage of planning to solve the online problem. Even a simple plan shows dramatic improvement. (based on simple statistical properties).}
\OK{Any offline plan can be used as input into OLIVE}}
\balance
    \bibliographystyle{IEEEtran}
    \bibliography{IEEEabrv,bib}

\end{document}